\documentclass[english]{article}
\usepackage[T1]{fontenc}
\usepackage[latin9]{inputenc}
\usepackage{amsmath}
\usepackage{cancel}
\usepackage{graphicx}

\makeatletter

\providecommand{\tabularnewline}{\\}

\pdfoutput=1
\usepackage{jheppub}
\usepackage{babel}
\usepackage{url}
\usepackage[numbers,sort&compress]{natbib}
\usepackage[subrefformat=parens,labelformat=parens]{subfig}
\newcommand{\zero}[1] {\cancelto{0}{#1}}
\newcommand{\tr}{\mathrm{Tr}\,}

\newcommand{\normord}[1]{{:}\!\mathrel{#1}\!{:}}
\newcommand{\ket}[1]{\left|#1\right\rangle}

\newcommand{\g}{\mathrm{g}}

\makeatother

\usepackage{babel}
\begin{document}
\author{Gaoli Chen} 
\affiliation{Institute for Fundamental Theory, Department of Physics, University of Florida, \\Gainesville, Florida 32611, USA} 
\emailAdd{gchen@ufl.edu}

\abstract{We provide a formalism to calculate the cubic interaction vertices of the stable string bit model, in which string bits have $s$ spin degrees of freedom but no space to move. With the vertices, we obtain a formula for one-loop self-energy, i.e., the $\mathcal{O}\left(1/N^{2}\right)$ correction to the energy spectrum. A rough analysis shows that, when the bit number $M$ is large, the ground state one-loop self-energy $\Delta E_{G}$ scale as $M^{5-s/4}$ for even $s$ and $M^{4-s/4}$ for odd $s$. Particularly, in $s=24$, we have $\Delta E_{G}\sim1/M$, which resembles the Poincar\'e invariant relation $P^{-}\sim1/P^{+}$ in $(1+1)$ dimensions. We calculate analytically the one-loop correction for the ground energies with $M=3$ and $s=1,\,2$. We then numerically confirm that the large $M$ behavior holds for $s\leq4$ cases.}

\title{Cubic interaction vertices and one-loop self-energy in the stable
string bit model}

\maketitle
\flushbottom

\section{Introduction}

In the string bit model \cite{Thorn:1991fv}, a string is a chain
comprised of pointlike entities called string bits. While the chain
is discretized, it behaves like a continuous string when the bit number
$M$ is large enough. 

The string bit model is an implementation of 't Hooft's idea of holography
\cite{'tHooft:1987,'tHooft:1990eb,'tHooft:1993gx}. In Lorentz invariant
theory, spacetime can be described by lightcone coordinates with transverse
dimensions $\mathbf{x}=\left(x^{2},\cdots,x^{D-1}\right)$ and the
`$\pm$' dimensions $x^{\pm}=\left(x^{0}\pm x^{1}\right)/\sqrt{2}$.
In the string bit model, the $x^{-}$ coordinate of string bits is
missing, and hence, the Lorentz invariance is not \textit{present
a priori}. String bits enjoy the dynamic of Galilean symmetry, under
which the $+$-component momentum $P^{+}=\left(P^{0}+P^{1}\right)/\sqrt{2}$
is identified as $mM$, where $m$ is the mass of one string bit.
When $M$ is large enough and $P^{+}$ is fixed, $P^{+}$ can be considered
as a continuous variable and its conjugate $x^{-}$ can be interpreted
as the missing coordinate. The Lorentz invariance can be therefore
regained and string theory emerges.

With 't Hooft's large $N$ limit \cite{'tHooft:1973jz,Thorn:1979gu},
the type II-B superstring was formulated in ref. \cite{Bergman:1995wh}
as a string bit model. In the model, a superstring bit creation operator,
which was an adjoint representation of $U\left(N\right)$ color group,
has up to $s$ spin indices and moves in transverse space. A more
drastic form of holography was studied in recent papers \cite{Sun:2014dga,Thorn:2014hia,Chen:2016hkz,Thorn:2015wli},
where string bits have no transverse coordinate and hence no space
to move. However, new compactified bosonic coordinates can be generated
from spin degrees of freedom of string bits. If suitable dynamics
is chosen, these spin degrees of freedom are converted to one-dimensional
spin waves, which then act as compactified bosonic coordinates. The
$1/N$ perturbation of the latter model was studied in ref. \cite{Thorn:2015wli},
where the cubic interaction vertices and their application to the
calculation of the one-loop self-energy were discussed.

Following the main idea of ref. \cite{Thorn:2015wli}, we continue
the work in the following way.

\begin{figure}
\begin{centering}
\includegraphics[width=0.5\textwidth]{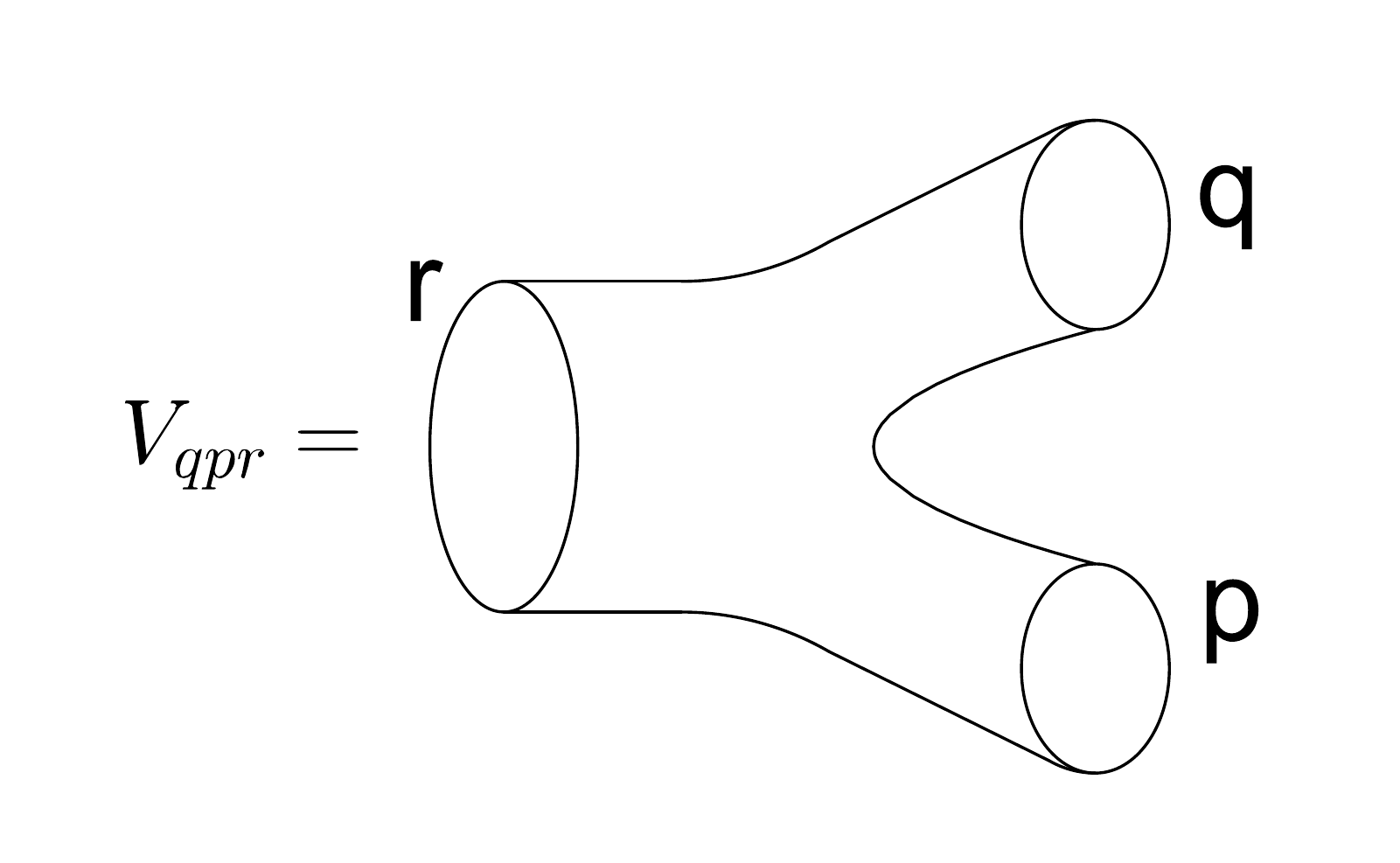}\includegraphics[width=0.5\textwidth]{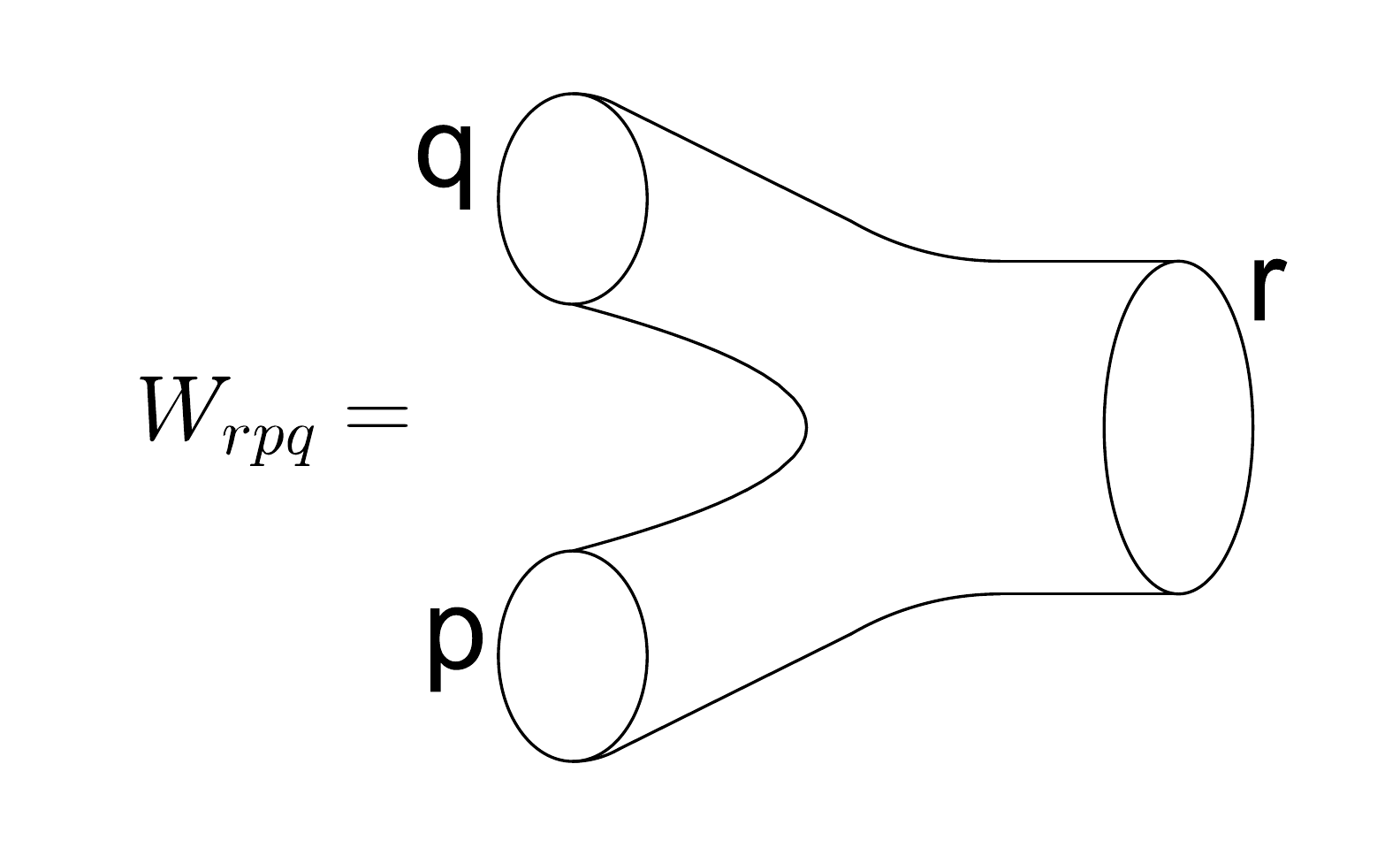}
\par\end{centering}
\caption{\label{fig:Vertices}The vertex $V_{qpr}$ is the amplitude of splitting
a large string $r$ into two small strings $p$ and $q$, while the
vertex $W_{rpq}$ is the amplitude of joining $p$ and $q$ into $r$. }
\end{figure}
\begin{itemize}
\item A more detailed study of the cubic interaction vertices is performed.
We present a systematic way to build conjugates of energy eigenfunctions,
determine the sign factors of the vertices, and (anti)symmetrize the
vertices, which are denoted as $V_{qpr}$ and $W_{rpq}$ and shown
as Figure \ref{fig:Vertices}, over the indices $p$ and $q$. We
then show that the interaction vertices can be calculated by finding
the vacuum expectation values of ladder operators. These are necessary
for the use of interaction vertices in our calculation of observables.
\item The calculation of the one-loop self-energy is improved, and its large
$M$ behavior for the ground states is analyzed. We assemble the ingredients
necessary to calculate the one-loop self-energy. The one-loop self-energies
of ground states, $\Delta E_{G}$, are studied, and their large $M$
behavior is analyzed. We calculate $\Delta E_{G}$ analytically for
the $M=3,\,s=1$ and $M=3,\,s=2$ cases. A qualitative analysis shows
that $\Delta E_{G}$ scales as $M^{5-s/4}$ for even $s$ and $M^{4-s/4}$
for odd $s$. The scaling behavior is consistent with Lorentz invariance
in $1+1$ dimensions when $s=24$, the critical Grassmann dimension,
and the protostring model \cite{Thorn:2015wli} emerges.
\item $\Delta E_{G}$ is determined numerically for higher $M$ and $s$.
We confirm the large $M$ behavior of $\Delta E_{G}$ for $s\leq4$.
We also verify that $\Delta E_{G}$ increases exponentially with respect
to $s$ when $M$ is fixed. We generalize the Hamiltonian of the model
by adding $\mathcal{O}\left(1/N\right)$ terms $s\xi\Delta H$ and
numerically show that, for the $s=2$ case, the Hamiltonian is bounded
from below with respect to $M$ only when $\xi\geq1$. Our analysis
suggests that this is true for all the even $s$ cases. The result
shows that the $s\xi\Delta H$ generalization is necessary for building
a physical string bit model. 
\end{itemize}
The rest of this paper is organized as follows. In section \ref{sec:Review},
we review some results of stable string bit models obtained by \cite{Thorn:2015wli}.
Specifically, we introduce the Hamiltonian of the model, solve for
the energy spectrum of the model at $N=\infty$, and summarize the
three chains overlap calculation. In section \ref{sec:Conjugate-Eigenfunction},
we provide a systematic approach to build conjugate eigenfunctions,
which will be used in the calculation of the $1/N$ expansion. In
section \ref{sec:interaction-vertex}, the cubic interaction vertices
are studied by $1/N$ perturbation. In section \ref{sec:EnergyCorrection},
we use the cubic interaction vertices to calculate one-loop self-energies.
Numerical results for the one-loop self-energy are analyzed in section
\ref{sec:Numerical-Results}. The main text is closed with a conclusion
section. Finally, several Appendixes are included for technical details.

\section{\label{sec:Review}Stable string bit model}

The purpose of this section is to review some results of stable string
bit models obtained in ref. \cite{Thorn:2015wli} and introduce useful
notations. These results are necessary for setting up the $1/N$ expansion
of the model. Meanwhile some modifications specific to this paper
are incorporated. To be clear, the modifications are as follows. In
Sec. \ref{subsec:Hamiltonian}, we add an $\mathcal{O}\left(1/N\right)$
term $\xi\Delta H$ to the Hamiltonian of the model. In Sec. \ref{subsec:Diagonalizing-Hamiltonian},
the diagonalization of the Hamiltonian at $N=\infty$ is done via
different intermediate variables.

\subsection{\label{subsec:Hamiltonian}Hamiltonian}

The superstring bit creation operator is 
\begin{equation}
\left(\bar{\phi}_{a_{1}\cdots a_{n}}\right)_{\alpha}^{\beta},\quad a_{i}=1,\cdots,s,\quad n=0,\cdots,s,\quad\alpha,\beta=1,\cdots,N.\label{eq:ham_creation_op}
\end{equation}
 where $a_{i}$ are totally antisymmetric spin indices and $\alpha$,
$\beta$ color indices of $U\left(N\right)$. $\bar{\phi}$ is bosonic
when $n$ is even and fermionic when $n$ is odd. In Fock space, a
closed string is represented by a color singlet trace operator acting
on the vacuum state, that is of the form $\tr\bar{\phi}\cdots\bar{\phi}\ket{0}$.
The number of $\bar{\phi}$ in the trace operator is the eigenvalue
of the bit number operator $M=\sum_{n}\frac{1}{n!}\tr\bar{\phi}_{a_{1}\cdots a_{n}}\phi_{a_{1}\cdots a_{n}}$. 

The Hamiltonian $H$ to be studied in this paper reads 
\begin{equation}
H=\sum_{i=1}^{5}H_{i}+s\xi\Delta H,\label{eq:ham_H_3}
\end{equation}
 where expressions of $H_{i}$ and $\Delta H$ are given in eqs. (\ref{eq:App_ham_subh})
and (\ref{eq:App_ham_deltah}). The $H_{i}$s make an $\mathcal{O}\left(1\right)$
contribution to $H$, while $\Delta H$ makes only $\mathcal{O}\left(1/N\right)$
contribution and hence does not affect the large $N$ limit. We note
that $H$ is a generalization of the $s=1$ Hamiltonian in refs. \cite{Sun:2014dga,Chen:2016hkz}.
The $H_{i}$ parts have been proposed in refs. \cite{Thorn:2014hia,Thorn:2015wli};
$\Delta H$ is the new term added by this paper and its derivation
is given in Appendix \ref{subsec:DeltaH}.

Let us now consider the action of $H$ on trace states space, which
is defined as follows. We introduce $s$ Grassmann coordinates $\theta^{a}$,
$a=1,\cdots,s$ and then define a superbit creation operator 
\[
\psi\left(\theta\right)=\sum_{k=0}^{s}\frac{1}{k!}\bar{\phi}_{c_{1}\cdots c_{k}}\theta^{c_{1}}\cdots\theta^{c_{k}}.
\]
and a single trace operator 
\[
T\left(\theta_{1},\cdots,\theta_{k}\right)=\tr\psi\left(\theta_{1}\right)\psi\left(\theta_{2}\right)\cdots\psi\left(\theta_{k}\right),
\]
 where $\theta_{i}$ are $s$-component Grassmann variables. The trace
states space, i.e., color singlet subspace of Fock space, is then
spanned by states like 
\[
T\left(\theta_{1},\cdots,\theta_{K}\right)T\left(\eta_{1},\cdots,\eta_{L}\right)\cdots\ket{0},
\]
 where $\ket{0}$ is the vacuum state. The action of each $H_{i}$
and $\Delta H$ on trace states is given in Appendix \ref{sec:app_hamiltonian}.
To summarize the results, let us define 
\begin{align}
\bar{h}_{kl} & =2\left(s-2\theta_{k}^{a}\frac{d}{d\theta_{k}^{a}}\right)+2\theta_{k}^{a}\frac{d}{d\theta_{l}^{a}}+2\theta_{l}^{a}\frac{d}{d\theta_{k}^{a}}-2i\theta_{k}^{a}\theta_{l}^{a}-2i\frac{d}{d\theta_{k}^{a}}\frac{d}{d\theta_{l}^{a}}+2s\xi-2s\delta_{k,l},\label{eq:ham_h_klbar}\\
\bar{h} & =\sum_{k=1}^{M}\left(\bar{h}_{k,k+1}-2s\xi\right).
\end{align}
Then the actions of $H$ on single and double trace states can be
written as\footnote{The actions of each $H_{i}$ on single and double trace states are
shown in Appendix \ref{sec:app_hamiltonian}.}
\begin{equation}
HT\left(\theta_{1},\cdots,\theta_{M}\right)\ket{0}=\bar{h}T\left(\theta_{1}\cdots\theta_{M}\right)\ket{0}+\frac{1}{N}\sum_{k=1}^{M}\sum_{l\neq k+1}\bar{h}_{kl}T\left(\theta_{l}\cdots\theta_{k}\right)T\left(\theta_{k+1}\cdots\theta_{l-1}\right)\ket{0}\label{eq:ham_Haction_single_trace}
\end{equation}
\begin{multline}
HT\left(\theta_{1}\cdots\theta_{K}\right)T\left(\eta_{1}\cdots\eta_{L}\right)\ket{0}=\left(\bar{h}_{\theta}+\bar{h}_{\eta}\right)T\left(\theta_{1}\cdots\theta_{K}\right)T\left(\eta_{1}\cdots\eta_{L}\right)\\
+\frac{1}{N}\sum_{k=1}^{K}\sum_{l=1}^{L}\bar{h}_{kl}T\left(\theta_{k+1}\cdots\theta_{k}\eta_{l}\cdots\eta_{l-1}\right)\ket{0}\\
+\frac{1}{N}\sum_{k=1}^{K}\sum_{l=1}^{L}\bar{h}_{lk}T\left(\theta_{k}\cdots\theta_{k-1}\eta_{l+1}\cdots\eta_{l}\right)\ket{0}+\frac{1}{N}\text{Fission Terms}.\label{eq:ham_Haction_double_trace}
\end{multline}
Note that in eqs. (\ref{eq:ham_Haction_double_trace}), the $-2s\delta_{kl}$
term of $\bar{h}_{kl}$ should be zero even if $k=l$, as they label
different variables.

While $\bar{h}_{kl}$ acts on the trace states, to solve for energy
eigenstates, it is helpful to convert $\bar{h}_{kl}$ to an equivalent
form acting on the wave function of an energy eigenstate at $N=\infty$.
The wave function $\psi_{r}$ is defined as follows. It follows from
eq. (\ref{eq:ham_Haction_single_trace}) that, at $N=\infty$, $H$
evolves single trace states to single trace states. Therefore, we
can express a single trace energy state as 
\begin{equation}
T_{r}\ket{0}=\int d^{s}\theta_{1}\cdots d^{s}\theta_{M}\,T\left(\theta_{1},\cdots,\theta_{M}\right)\psi_{r}\left(\theta_{1},\dots,\theta_{M}\right)\ket{0},\label{eq:ham_trace_energystate}
\end{equation}
 where $\psi_{r}$ is the wave function. Since $T\left(\theta_{1},\cdots,\theta_{M}\right)$
is invariant under the cyclic permutation $\theta_{i}\to\theta_{i+1}$,
we can constrain $\psi_{r}$ by 
\begin{equation}
\psi_{r}\left(\theta_{1},\cdots,\theta_{M}\right)=\left(-\right)^{s\left(M-1\right)}\psi_{r}\left(\theta_{2},\cdots,\theta_{M},\theta_{1}\right)\label{eq:ham_cyclicconstrtaint}
\end{equation}
 without loss of generality. The sign factor follows from the fact
that the measure $d^{s}\theta_{1}\cdots d^{s}\theta_{M}$ is changed
by a factor $\left(-\right)^{s\left(M-1\right)}$ under the cyclic
transformation $\theta_{i}\to\theta_{i+1}$. Now, the action of $\bar{h}_{kl}$
on $T_{r}\ket{0}$ is
\begin{equation}
\bar{h}_{kl}T_{r}\ket{0}=\int d\theta\,\bar{h}_{kl}T\left(\theta\right)\psi_{r}\left(\theta\right)\ket{0}=\int d\theta\,T\left(\theta\right)h_{kl}\psi_{r}\left(\theta\right)\ket{0}\label{eq:ham_intbyparts}
\end{equation}
 where we have performed an integration by parts in the last step
and
\begin{equation}
h_{kl}=-2\left(s-2\theta_{k}^{a}\frac{d}{d\theta_{k}^{a}}\right)-2\theta_{k}^{a}\frac{d}{d\theta_{l}^{a}}-2\theta_{l}^{a}\frac{d}{d\theta_{k}^{a}}-2i\theta_{k}^{a}\theta_{l}^{a}-2i\frac{d}{d\theta_{k}^{a}}\frac{d}{d\theta_{l}^{a}}+2s\xi+2s\delta_{k,l}.\label{eq:ham_h_kl}
\end{equation}
Note that, in the derivation of $h_{kl}$, the $k=l$ case needs special
treatment. Likewise, the action of $\bar{h}$ on $T\left(\theta\right)$
is equivalent to the action on $\psi_{r}\left(\theta\right)$ by 
\begin{equation}
h=\sum_{k=1}^{M}\left(h_{k,k+1}-2s\xi\right).\label{eq:ham_h}
\end{equation}

\subsection{\label{subsec:Diagonalizing-Hamiltonian}Diagonalizing Hamiltonian
at $N=\infty$}

Now let us solve for the energy spectrum of the model at $N=\infty$.
A single trace energy eigenstate is determined by an eigenfunction
$\psi_{r}$ satisfying the equation 
\begin{equation}
h\psi_{r}\left(\theta_{1},\dots,\theta_{M}\right)=E_{r}\psi_{r}\left(\theta_{1},\dots,\theta_{M}\right).\label{eq:dia_eigenequation}
\end{equation}
To solve the eigenvalue problem eq. (\ref{eq:dia_eigenequation}),
we need to find the lowering and raising eigenoperators of $h$. This
has been done by ref. \cite{Thorn:2015wli}. Here, we repeat the procedure
with different sets of intermediate variables.

From (\ref{eq:ham_h_kl}), we see that each term of $h$ contains
only variables or derivatives of the same $\theta^{a}$. It implies
the variables can be separated and we only need to solve the equation
of one variable. We therefore drop the spin index $a$ in the following
calculation.

We introduce Fourier transforms \cite{Chen:2016hkz,Sun:2014dga}\begin{subequations}\label{dia_fourier}
\begin{align}
\alpha_{n} & =\frac{1}{\sqrt{M}}\sum_{k=1}^{M}\theta_{k}e^{-2\pi ikn/M},\quad\beta_{n}=\frac{1}{\sqrt{M}}\sum_{k=1}^{M}\frac{d}{d\theta_{k}}e^{-2\pi ikn/M},\label{eq:dia_fourier:1}\\
\theta_{k} & =\frac{1}{\sqrt{M}}\sum_{n=0}^{M-1}\alpha_{n}e^{2\pi ikn/M},\quad\frac{d}{d\theta_{k}}=\frac{1}{\sqrt{M}}\sum_{n=0}^{M-1}\beta_{n}e^{2\pi ikn/M},\label{eq:dia_fourier:2}\\
n & =0,\dots M-1,\quad k=1,\dots M.\nonumber 
\end{align}
\end{subequations}which satisfy

\begin{equation}
\left\{ \alpha_{n},\beta_{m}\right\} =\delta_{m+n,M}+\delta_{m,0}\delta_{n,0}.\label{eq:dia_alpha_beta_anticom}
\end{equation}
In ref. \cite{Thorn:2015wli}, instead of $\theta_{k}$ and $\frac{d}{d\theta_{k}}$,
the diagonalization was done via the Grassmann variables $S_{k}=\theta_{k}+\frac{d}{d\theta_{k}}$,
$\tilde{S}_{k}=i\left(\theta_{k}+\frac{d}{d\theta_{k}}\right)$, and
their Fourier transforms. Such different choices should not affect
the eigenoperators and the energy spectrum. 

The Hermiticity of the Hamiltonian implies that $\theta_{k}^{\dagger}=\frac{d}{d\theta_{k}},\,\frac{d}{d\theta_{k}}^{\dagger}=\theta_{k},$
from which it follows that 
\[
\alpha_{n}^{\dagger}=\beta_{M-n},\quad\beta_{n}^{\dagger}=\alpha_{M-n},\quad\left\{ \alpha_{n},\alpha_{m}^{\dagger}\right\} =\left\{ \beta_{n},\beta_{m}^{\dagger}\right\} =\delta_{m,n},\quad0\leq n,m\leq M-1.
\]
 We now express $h$ in terms of $\alpha_{n}$ and $\beta_{n}$ as
\begin{equation}
h=2\sum_{n=1}^{M-1}\left[\left(\alpha_{n}\alpha_{M-n}+\beta_{n}\beta_{M-n}\right)\sin\frac{2n\pi}{M}+2\left(1-\cos\frac{2n\pi}{M}\right)\left(\alpha_{n}\beta_{M-n}+\alpha_{M-n}\beta_{n}\right)\right]-2M,\label{eq:dia_h_ab}
\end{equation}
and seek for eigenoperators of $h$, 
\begin{equation}
F_{k}=r_{k}\alpha_{k}+\beta_{k},\quad\left[h,F_{k}\right]=\epsilon_{k}F_{k},\label{eq:dia_operator_eq}
\end{equation}
 where $r_{k}$ and $\epsilon_{k}$ are constants. Substituting (\ref{eq:dia_h_ab})
into (\ref{eq:dia_operator_eq}) yields
\[
\epsilon_{k}^{\pm}=\pm8\sin\frac{k\pi}{M},\quad r_{k}^{\pm}=\tan\frac{k\pi}{M}\pm\sec\frac{k\pi}{M}.
\]
We then normalize the coefficients of $F_{k}$ to obtain the lowering
and raising operators for $k\geq1$,\begin{subequations}\label{eq:dia_F}
\begin{equation}
F_{k}=s_{k}\alpha_{k}+c_{k}\beta_{k},\quad\bar{F}_{k}=c_{k}\alpha_{k}-s_{k}\beta_{k},\quad k=1,\cdots,M-1,\label{eq:dia_F:1}
\end{equation}
where $c_{k}=\cos\left(\frac{\pi}{4}-\frac{k\pi}{2M}\right)$ and
$s_{k}=\sin\left(\frac{\pi}{4}-\frac{k\pi}{2M}\right)$. It follows
from (\ref{eq:dia_F:1}) that 
\begin{equation}
F_{k}^{\dagger}=\bar{F}_{M-k}=c_{k}\alpha_{M-k}+s_{k}\beta_{M-k},\quad1\leq k\leq M-1,\label{eq:dia_F:2}
\end{equation}
The zero modes need special treatment: 
\begin{equation}
F_{0}=F_{M}=e^{i\pi/4}\beta_{0},\quad F_{0}^{\dagger}=\bar{F}_{0}=\bar{F}_{M}=e^{-i\pi/4}\alpha_{0}.\label{eq:dia_F:3}
\end{equation}
\end{subequations}The phase factors are chosen so that the expression
of $h_{kl}$ in terms of eigenoperators will have a simple form; see
eq. (\ref{eq:cub_h_kl}). A direct calculation shows that the eigenoperators
satisfy the following anticommutation relations
\begin{equation}
\left\{ F_{k},F_{l}\right\} =\left\{ F_{k}^{\dagger},F_{l}^{\dagger}\right\} =0,\quad\left\{ F_{k},F_{l}^{\dagger}\right\} =\delta_{kl},\quad0\leq k,l\leq M-1.\label{eq:dia_F_anticommutation}
\end{equation}

To obtain the energy spectrum, we need to find the ground energy $E_{G}$
and the ground eigenfunction $\psi_{G}$, which is annihilated by
all the lowering operators. Since the zero mode does not change energy
eigenvalues, there are degeneracies in ground state. To eliminate
the ambiguity, we require the ground eigenfunction to be annihilated
by the zero mode $F_{0}$ as well. The ground eigenfunction can be
\cite{Chen:2016hkz} 
\begin{equation}
\psi_{G}^{s=1}=\prod_{k=1}^{\left\lfloor \left(M-1\right)/2\right\rfloor }\left(c_{k}-s_{k}\alpha_{M-k}\alpha_{k}\right),\label{eq:dia_ground_eigenfunction}
\end{equation}
 where $\left\lfloor \left(M-1\right)/2\right\rfloor $ indicates
the integral part of $\left(M-1\right)/2$. To verify $F_{m}\psi_{G}^{s=1}=0$,
one only needs to check that
\begin{align}
F_{k}\left(c_{k}-s_{k}\alpha_{M-k}\alpha_{k}\right) & =F_{M-k}\left(c_{k}-s_{k}\alpha_{M-k}\alpha_{k}\right)=0,\quad1\leq k\leq M-1,\\
\left[F_{k},c_{l}-s_{l}\alpha_{M-l}\alpha_{l}\right] & =0,\quad k\neq l,\,k\neq M-l.
\end{align}
Acting $h$ on the ground eigenfunction, we obtain the ground energy
\begin{equation}
E_{G}^{s=1}=-4\sum_{k=1}^{M-1}\sin\frac{k\pi}{M}=-4\cot\frac{\pi}{2M}=-\frac{8M}{\pi}+\frac{2\pi}{3M}+\mathcal{O}\left(M^{-3}\right).\label{eq:diag_ground_energy}
\end{equation}

We can now build general eigenfunctions for arbitrary $s$ case. The
ground eigenfunction and energy are 
\begin{equation}
\psi_{G}=\psi_{G}^{\left(1\right)}\psi_{G}^{\left(2\right)}\cdots\psi_{G}^{\left(s\right)},\quad E_{G}=-4s\cot\frac{\pi}{2M},\label{eq:dia_ground_state_s}
\end{equation}
 where each $\psi_{G}^{\left(a\right)}$ has the form of (\ref{eq:dia_ground_eigenfunction}).
A general energy eigenfunction $\psi_{r}$ and its corresponding energy
can be written as\begin{subequations}
\begin{align}
\psi_{r} & =\left(F_{r_{1,1}}^{\left(1\right)}F_{r_{1,2}}^{\left(1\right)}\cdots F_{r_{2,1}}^{\left(2\right)}F_{r_{2,2}}^{\left(2\right)}\cdots F_{r_{s,1}}^{\left(s\right)}F_{r_{s,2}}^{\left(s\right)}\cdots\right)^{\dagger}\psi_{G}\equiv F_{\left\{ r\right\} }^{\dagger}\psi_{G},\label{eq:dia_general_eigenfunction_s}\\
E_{r} & =-4s\cot\frac{\pi}{2M}+8\sum_{a,i}\sin\frac{r_{a,k}\pi}{M},\label{eq:dia_general_energy_s}
\end{align}
\end{subequations} where we have defined $F_{\left\{ r\right\} }$
as a string of eigenoperators and we choose $0\leq r_{a,1}<r_{a,2}<\cdots\leq M-1$
as a convention. To build a physical state, the modes $r_{a,i}$ (\ref{eq:dia_general_eigenfunction_s})
need to satisfy the cyclic constraint (\ref{eq:ham_cyclicconstrtaint}).
Under the cyclic permutation $\theta_{k}^{a}\to\theta_{k+1}^{a}$,
$F_{k}^{a\dagger}$ transforms as $F_{k}^{a\dagger}\to e^{-2ik\pi/M}F_{k}^{a\dagger}$.
It then follows from eq. (\ref{eq:ham_cyclicconstrtaint}) that the
modes must satisfy
\begin{equation}
\sum_{a,k}r_{a,k}=\begin{cases}
nM & \quad\text{for even }s\left(M-1\right)\\
\left(n+\frac{1}{2}\right)M & \quad\text{for odd }s\left(M-1\right)
\end{cases},\quad n=0,\,1,\,2,\cdots\label{eq:dia_cyclic_constraint_s}
\end{equation}
Since the zero modes do not change the energy, the ground energy eigenstate
has at least $2^{s}$ degeneracies. This is the consequence of $H$
commuting with supersymmetry operators $Q^{a}$, as defined in eq.
(\ref{eq:App_ham_Q}). The constraint (\ref{eq:dia_cyclic_constraint_s})
has a profound impact on the energy spectrum of the model. When $s$
is even, all the ground states are allowed by (\ref{eq:dia_cyclic_constraint_s})
and are hence physical. But when $s$ is odd, the ground state is
allowed only when $M$ is odd. It then follows that the lowest single
trace state for even $M$ is the one corresponding to $F_{M/2}^{a\dagger}\psi_{G}$.

\subsection{\label{subsec:Three-Chain-overlap}Three chains overlap}

We have constructed the energy eigenfunctions for $N=\infty$. To
obtain the $1/N$ expansion results, we also need to calculate the
overlap among three chains: one large chain of $M$ bits and two small
chains of $K$ bits and $L=M-K$ bits. The calculation can be done
by establishing the relation among the eigenoperators of large chain
and two small chains. Here, we recap the results of ref. \cite{Thorn:2015wli}.

Let us only consider the $s=1$ case. Let $F_{m}^{\left(K\right)}$
and $F_{n}^{\left(L\right)}$ be lowering operators of $L$-bit and
$K$-bit chains. Define a set of operators\begin{subequations}\label{eq:overlap_f}
\begin{align}
f_{0} & =F_{0}^{\left(L\right)}\sqrt{\frac{L}{M}}+F_{0}^{\left(K\right)}\sqrt{\frac{K}{M}},\label{eq:overlap_f:1}\\
f_{n} & =F_{l}^{\left(L\right)},\quad1\leq n\leq L-1,\label{eq:overlap_f:2}\\
f_{n+L} & =F_{k}^{\left(K\right)},\quad1\leq n\leq K-1,\label{eq:overlap_f:3}\\
f_{M-1} & =e^{-i\pi/4}\left(F_{0}^{\left(L\right)}\sqrt{\frac{K}{M}}-F_{0}^{\left(K\right)}\sqrt{\frac{L}{M}}\right),\label{eq:overlap_f:4}
\end{align}
\end{subequations} which satisfy the anticommutation relationship
$\left\{ f_{n},f_{m}\right\} =\left\{ f_{n}^{\dagger},f_{m}^{\dagger}\right\} =0$
and $\left\{ f_{n},f_{m}^{\dagger}\right\} =\delta_{nm}$. Note that
$f_{0}$ equals $F_{0}$ of the large chain \cite{Thorn:2015wli}.
We then express the large chain operators in terms of $f$ and $f^{\dagger}$
as 
\begin{equation}
F_{m}=\sum_{n=0}^{M-1}\left(f_{n}C_{mn}+f_{n}^{\dagger}S_{mn}\right),\quad0\leq m\leq M-1.\label{eq:overlap_Ff_relation}
\end{equation}
The anticommutation relation among $F_{m}$ and $F_{m}^{\dagger}$
requires 
\begin{equation}
CS^{T}+SC^{T}=0,\quad CC^{\dagger}+SS^{\dagger}=I.\label{eq:overlap_CS_relation}
\end{equation}
The matrix elements of $C$ and $S$ are given by
\[
C_{0n}=C_{n0}=\delta_{0,n},\quad S_{0,n}=S_{n,0}=0,\quad0\leq n<M
\]
 and \cite{Thorn:2015wli}\begin{subequations}\label{eq:overlap_CS}
\begin{align}
C_{mn} & =-\frac{1}{\sqrt{ML}}\frac{1-e^{-2\pi imL/M}}{1-e^{-2\pi i\left(n/L-m/M\right)}}\cos\left(\frac{n\pi}{2L}-\frac{m\pi}{2M}\right),\quad1\leq n<L\label{eq:overlap_CS:1}\\
C_{m,L+n-1} & =\frac{1}{\sqrt{MK}}\frac{1-e^{-2\pi imL/M}}{1-e^{-2\pi i\left(n/K-m/M\right)}}\cos\left(\frac{n\pi}{2K}-\frac{m\pi}{2M}\right),\quad1\leq n<K\label{eq:overlap_CS:2}\\
C_{m,M-1} & =-\frac{1}{\sqrt{LK}}\frac{1-e^{-2\pi imL/M}}{1-e^{2i\pi m/M}}\cos\left(\frac{m\pi}{2M}-\frac{\pi}{4}\right),\label{eq:overlap_CS:3}\\
S_{mn} & =-\frac{1}{\sqrt{ML}}\frac{1-e^{-2\pi imL/M}}{1-e^{2\pi i\left(n/L+m/M\right)}}\cos\left(\frac{n\pi}{2L}+\frac{m\pi}{2M}\right),\quad1\leq n<L\label{eq:overlap_CS:4}\\
S_{m,L+n-1} & =\frac{1}{\sqrt{MK}}\frac{1-e^{-2\pi imL/M}}{1-e^{2\pi i\left(n/K+m/M\right)}}\cos\left(\frac{n\pi}{2K}+\frac{m\pi}{2M}\right),\quad1\leq n<K\label{eq:overlap_CS:5}\\
S_{m,M-1} & =-\frac{1}{\sqrt{LK}}\frac{1-e^{-2\pi imL/M}}{1-e^{2\pi im/M}}\cos\left(\frac{m\pi}{2M}+\frac{\pi}{4}\right),\label{eq:overlap_CS:6}
\end{align}
\end{subequations}where $1\leq m\leq M-1$ in eqs. (\ref{eq:overlap_CS}).
When $M$ is large, the determinate of $C$ can be approximated as
\cite{Thorn:2015wli}
\begin{equation}
\det CC^{\dagger}\sim\frac{0.9290}{\left(KLM\right)^{1/6}}\left(\frac{L}{M}\right)^{\left(M/K-L/M\right)/3-2/3}\left(\frac{K}{M}\right)^{\left(M/L-K/M\right)/3-2/3}.\label{eq:overlap_detC}
\end{equation}

We then express the ground eigenfunction of the large chain as
\begin{equation}
\psi_{G}^{\left(M\right)}=\exp\left(\frac{1}{2}\sum_{kl}f_{k}^{\dagger}D_{kl}f_{l}^{\dagger}\right)\psi_{G}^{\left(K\right)}\psi_{G}^{\left(L\right)}\left[\det\left(I+DD^{\dagger}\right)\right]^{-1/4},\label{eq:overlap_psi_M}
\end{equation}
 where $\psi_{G}^{\left(K\right)}$ and $\psi_{G}^{\left(L\right)}$
are ground eigenfunction for two small chains. The constraints $F_{m}\psi_{G}=0$
imply 
\begin{equation}
C_{mn}D_{nl}+S_{ml}=0.\label{eq:overlap_CMS}
\end{equation}

From the above construction, it is clear that the first rows and columns
of the matrices $C$, $S$, and $D$ are trivial. One can therefore
write them as $C=\begin{pmatrix}1\end{pmatrix}\oplus C^{\prime}$,
$S=\begin{pmatrix}0\end{pmatrix}\oplus S^{\prime}$, and $D=\begin{pmatrix}0\end{pmatrix}\oplus D^{\prime}$
where $C^{\prime}$, $S^{\prime}$, and $D^{\prime}$ are nontrivial
matrices of dimension $\left(M-1\right)\times\left(M-1\right)$. 

With (\ref{eq:overlap_CS_relation}) and (\ref{eq:overlap_CMS}),
we can simplify (\ref{eq:overlap_psi_M}): 
\[
\det\left(I+DD^{\dagger}\right)=\frac{\det\left[C\left(I+DD^{\dagger}\right)C^{\dagger}\right]}{\det\left[CC^{\dagger}\right]}=\frac{\det\left[CC^{\dagger}+SS^{\dagger}\right]}{\det\left(CC^{\dagger}\right)}=\left|\det C\right|^{-2},
\]
 
\begin{equation}
\psi_{G}^{\left(M\right)}=\left|\det C\right|^{1/2}\exp\left(\frac{1}{2}\sum_{kl}f_{k}^{\dagger}D_{kl}f_{l}^{\dagger}\right)\psi_{G}^{\left(K\right)}\psi_{G}^{\left(L\right)}.\label{eq:overlap_psi_M1}
\end{equation}

\section{\label{sec:Conjugate-Eigenfunction}Conjugate eigenfunction}

We have built energy eigenfunctions of the model at $N=\infty$ in
Sec. \ref{subsec:Diagonalizing-Hamiltonian}. To calculate $1/N$
expansion results, we also need to find functions that conjugate to
the energy eigenfunctions. For convenience, we call these functions
conjugate eigenfunctions. In this section, we will construct conjugate
eigenfunctions systematically. 

A conjugate eigenfunction $\bar{\psi}_{r}$ is a function of $\theta_{i}$
that satisfies the normalization condition \cite{Thorn:2015wli} 
\begin{equation}
\int d^{s}\theta_{1}\cdots d^{s}\theta_{M}\,\bar{\psi}_{r}\left(\theta_{1},\cdots,\theta_{M}\right)\psi_{s}\left(\theta_{1},\cdots,\theta_{M}\right)=\delta_{rs}\label{eq:conj_normalization}
\end{equation}
 and the completeness relation 
\begin{equation}
\sum_{r}\psi_{r}\left(\theta_{1},\cdots,\theta_{M}\right)\bar{\psi}_{r}\left(\eta_{1},\cdots,\eta_{M}\right)=\tilde{\delta}\left(\theta-\eta\right),\label{eq:conj_completeness}
\end{equation}
where the delta function $\tilde{\delta}\left(\theta-\eta\right)$
is understood to be symmetrized under cyclic constraint like (\ref{eq:ham_cyclicconstrtaint}).\footnote{To be specific, it means that 
\[
\int d^{M}\theta\,f\left(\theta_{1},\cdots,\theta_{M}\right)\tilde{\delta}\left(\theta-\eta\right)=\frac{1}{M}\sum_{k=0}^{M-1}\left(-\right)^{ks\left(M-1\right)}f\left(\eta_{k+1},\cdots,\eta_{k+M}\right).
\]
} We stress that, once there is a complete set of $\bar{\psi}_{r}$
and $\psi_{r}$ fulfilling the normalization condition, the completeness
relation is satisfied automatically. 

To construct $\bar{\psi}_{r}$ explicitly, it is convenient to define
operators $\tilde{F}_{k}^{\pm}$ as conjugate to $F_{k}$ under integration
by parts, 
\begin{equation}
\int d^{M}\theta\,\psi\left(\theta\right)\left[F_{k}\chi\left(\theta\right)\right]=\int d^{M}\theta\,\left[\tilde{F}_{k}^{\pm}\psi\left(\theta\right)\right]\chi\left(\theta\right),\label{eq:conj_Ftilde1}
\end{equation}
where the $+$ superscript is chosen if $\psi\left(\theta\right)$
is Grassmann even and $-$ is chosen otherwise. It then follows from
eqs. (\ref{eq:dia_F}) that 
\begin{equation}
\tilde{F}_{0}^{\pm}=\mp e^{i\pi/4}\beta_{0},\quad\tilde{F}_{k}^{\pm}=\pm\left(s_{k}\alpha_{k}-c_{k}\beta_{k}\right),\quad1\leq k\leq M-1,
\end{equation}
\begin{equation}
\tilde{F}_{0}^{\dagger\pm}=\pm e^{-i\pi/4}\alpha_{0},\quad\tilde{F}_{k}^{\dagger\pm}=\pm\left(c_{k}\alpha_{M-k}-s_{k}\beta_{M-k}\right),\quad1\leq k\leq M-1.\label{eq:conj_F_dagger_tilder}
\end{equation}
In the remainder of this paper, we may suppress the superscript $\pm$
if there is no danger of ambiguity.

In the $s=1$ case, we claim that the conjugate to the ground eigenfunction
$\psi_{G}^{s=1}$ is\begin{subequations}\label{eq:conj_psibarG}
\begin{align}
\bar{\psi}_{G}^{s=1} & =\left(-i\right)^{\left\lfloor M/2\right\rfloor }\alpha_{0}\prod_{i=1}^{\left\lfloor M/2\right\rfloor }\left(-s_{i}+c_{i}\alpha_{M-i}\alpha_{i}\right)\quad\text{for odd }M\label{eq:conj_psibarG:1}\\
\bar{\psi}_{G}^{s=1} & =\left(-i\right)^{M/2+1}\prod_{i=1}^{\left\lfloor \left(M-1\right)/2\right\rfloor }\left(-s_{i}+c_{i}\alpha_{M-i}\alpha_{i}\right)\alpha_{0}\alpha_{M/2}\quad\text{for even }M.\label{eq:conj_psibarG:2}
\end{align}
\end{subequations} In Appendix \ref{sec:App_normal_psibar}, we verify
that $\bar{\psi}_{G}^{s=1}$ satisfies the normalization condition
(\ref{eq:conj_normalization}). The function conjugate to the general
eigenfunction (\ref{eq:dia_general_eigenfunction_s}) can be built
by acting on $\bar{\psi}_{G}$ with a string of $\tilde{F}_{k}^{\left(a\right)}$
as\begin{subequations}
\begin{align}
\bar{\psi}_{r} & =\tilde{F}_{r_{1,1}}^{\left(1\right)}\tilde{F}_{r_{1,2}}^{\left(1\right)}\cdots\tilde{F}_{r_{2,1}}^{\left(2\right)}\tilde{F}_{r_{2,2}}^{\left(2\right)}\cdots\tilde{F}_{r_{s,1}}^{\left(s\right)}\tilde{F}_{r_{s,2}}^{\left(s\right)}\cdots\bar{\psi}_{G}\equiv\tilde{F}_{\left\{ r\right\} }\bar{\psi}_{G},\label{eq:conj_psibar_general}\\
\bar{\psi}_{G} & =\left(-\right)^{s\left(s-1\right)M\left(M-1\right)/4}\psi_{G}^{\left(s\right)}\psi_{G}^{\left(s-1\right)}\cdots\psi_{G}^{\left(1\right)},
\end{align}
\end{subequations}where all the $\tilde{F}$s pick $\tilde{F}^{+}$
if $\bar{\psi}_{G}$ is Grassmann even and $\tilde{F}^{-}$ otherwise.
The normalization condition (\ref{eq:conj_normalization}) can be
easily verified, 
\begin{align*}
\int d^{M}\theta\,\bar{\psi}_{r}\psi_{r} & =\int d^{M}\theta\,\tilde{F}_{\left\{ r\right\} }\bar{\psi}_{G}F_{\left\{ r\right\} }^{\dagger}\psi_{G}\\
 & =\int d^{M}\theta\,\bar{\psi}_{G}F_{\left\{ r\right\} }F_{\left\{ r\right\} }^{\dagger}\psi_{G}\\
 & =\int d^{M}\theta\,\bar{\psi}_{G}\psi_{G}=1,
\end{align*}
 where we used (\ref{eq:conj_Ftilde1}) in the second equality and
(\ref{eq:dia_F_anticommutation}) in the third equality. In the last
equality, the sign factor of $\bar{\psi}_{G}$ cancels the sign introduced
by the rearrangement of the measure from $d^{s}\theta_{1}\cdots d^{s}\theta_{M}$
to $\left(\prod_{i=1}^{M}d\theta_{i}^{\left(1\right)}\right)\cdots\left(\prod_{i=1}^{M}d\theta_{i}^{\left(s\right)}\right)$.

By analogy with (\ref{eq:overlap_psi_M}), for the $s=1$ case, the
overlap of conjugate eigenfunctions among the large chain and two
small chains is given by
\begin{equation}
\bar{\psi}_{G}^{\left(M\right)}=\left|\det C\right|^{1/2}\exp\left(\frac{1}{2}\sum_{ij}\tilde{f}_{i}D_{ij}^{\dagger}\tilde{f}_{j}\right)\bar{\psi}_{G}^{\left(K\right)}\bar{\psi}_{G}^{\left(L\right)},\label{eq:conj_psibar_M}
\end{equation}
 where $\tilde{f}$ picks $\tilde{f}^{+}$ if $M$ is even and $\tilde{f}^{-}$
if $M$ is odd and all the notations follow the ones of Sec. \ref{subsec:Three-Chain-overlap}.

Let us conclude this section by discussing the grading of energy eigenstates
and eigenfunctions. We define 
\[
g_{r}\equiv\g\left(T_{r}\right)=\text{grading of }T_{r}.
\]
Now, we can write the trace operator $T\left(\theta\right)$ as a
linear combination of $\bar{\psi}_{r}$. Let $T\left(\theta\right)=\sum_{t}X_{t}\bar{\psi}_{t}\left(\theta\right)$,
where $X_{t}$ is independent of $\theta$; then 
\begin{align*}
T_{r} & =\int d^{s}\theta_{1}\cdots d^{s}\theta_{M}\,T\left(\theta\right)\psi_{r}\left(\theta\right)\\
 & =\sum_{t}\left(-\right)^{sM\g\left(X_{t}\right)}X_{t}\int d^{s}\theta_{1}\cdots d^{s}\theta_{M}\,\bar{\psi}_{t}\left(\theta\right)\psi_{r}\left(\theta\right)\\
 & =\left(-\right)^{sM\g\left(X_{r}\right)}X_{r},
\end{align*}
 where the sign factor comes from the commutation of the measure and
$X_{t}$. It implies that $X_{r}$ differs from $T_{r}$ only by a
sign factor. So, we have $\g\left(X_{r}\right)=\g\left(T_{r}\right)$
and 
\begin{equation}
T\left(\theta\right)=\sum_{r}\left(-\right)^{sMg_{r}}T_{r}\bar{\psi}_{r}\left(\theta\right).\label{eq:conj_completeness_1}
\end{equation}
Finally, from eqs. (\ref{eq:conj_completeness_1}), (\ref{eq:conj_normalization}),
and (\ref{eq:dia_ground_state_s}), we obtain the gradings (modulo
2) of functions and operators as Table \ref{tab:gradings}. These
results will be used in the next section.

\begin{table}
\begin{centering}
\begin{tabular}{|c|c|c|c|c|c|}
\hline 
$T_{r}$ & $\psi_{r}$ & $\bar{\psi}_{r}$ & $\psi_{G}$ & $\bar{\psi}_{G}$ & $F_{\left\{ r\right\} }$\tabularnewline
\hline 
\hline 
$g_{r}$ & $g_{r}-sM$ & $g_{r}$ & $0$ & $sM$ & $g_{r}-sM$\tabularnewline
\hline 
\end{tabular}
\par\end{centering}
\caption{\label{tab:gradings}Gradings of functions and operators.}

\end{table}

\section{\label{sec:interaction-vertex}Cubic interaction vertices}

Let $T_{p}\ket{0}$, $T_{q}\ket{0}$, and $T_{r}\ket{0}$ be energy
eigenstates of strings with $K$, $L$, and $M=K+L$ bits respectively;
then the interaction vertices $V_{qpr}$ and $W_{rpq}$ are defined
as\cite{Thorn:2015wli}\begin{subequations}\label{eq:cub_Vertices_Def}
\begin{align}
HT_{r}\ket{0} & =E_{r}T_{r}\ket{0}+\frac{1}{N}\sum_{K=1}^{M-1}\sum_{p,q}T_{p}T_{q}\ket{0}V_{qpr},\label{eq:cub_Vertices_DefV}\\
HT_{p}T_{q}\ket{0} & =\left(E_{p}+E_{q}\right)T_{p}T_{q}\ket{0}+\frac{1}{N}\sum_{r}T_{r}\ket{0}W_{rpq}+\cdots.\label{eq:cub_Vertices_DefW}
\end{align}
\end{subequations} The vertex $V_{qpr}$ represents the amplitude
of breaking one large string into two small strings and the vertex
$W_{rpq}$ represents the amplitude of joining two small strings into
one large string. Without loss of generality, we can (anti)symmetrize
the vertices over indices $p$ and $q$ as 
\begin{equation}
V_{pqr}=\left(-\right)^{g_{p}g_{q}}V_{qpr},\quad W_{rqp}=\left(-\right)^{g_{p}g_{q}}W_{rpq}.\label{eq:cub_Vertices_Sym}
\end{equation}
In this section, we shall find that\begin{subequations}\label{eq:cub_Vertices}
\begin{align}
V_{qpr} & =M\left|\det C\right|^{s/2}\prod_{a=1}^{s}\left\langle F_{\left\{ qp\right\} }^{a},F_{\left\{ r\right\} }^{a\dagger}\right\rangle _{V},\label{eq:cub_VertexV}\\
W_{rpq} & =KL\left|\det C\right|^{s/2}\prod_{a=1}^{s}\left\langle F_{\left\{ qp\right\} }^{a},F_{\left\{ r\right\} }^{a\dagger}\right\rangle _{W}^{\dagger}.\label{eq:cub_VertexW}
\end{align}
\end{subequations}Several notations are used in (\ref{eq:cub_Vertices})
for convenience. $F_{\left\{ qp\right\} }^{a}\equiv F_{\left\{ q\right\} }^{a}F_{\left\{ p\right\} }^{a}$
and the superscript $a$ indicates that only operators of spin index
$a$ are involved. The brackets $\left\langle \cdot,\cdot\right\rangle _{V,W}$
stand for vacuum expectation values of operators\begin{subequations}
\begin{align}
\left\langle F_{\left\{ qp\right\} }^{a},F_{\left\{ r\right\} }^{a\dagger}\right\rangle _{V,W} & \equiv\left\langle F_{\left\{ qp\right\} }^{a}h_{\left(V,W\right)}^{a}F_{\left\{ r\right\} }^{a\dagger}\exp\left(\frac{1}{2}f_{k}^{a\dagger}D_{kl}f_{l}^{a\dagger}\right)\right\rangle ,\label{eq:cub_vac_VW}\\
h_{V}^{a} & \equiv\frac{1}{2}\left(h_{K,1}^{a}+h_{M,K+1}^{a}\right),\quad h_{W}^{a}=h_{K,K+1}^{a}+h_{M,1}^{a}\label{eq:cub_hVW}
\end{align}
\end{subequations}where the matrix $D$ and operators $f_{k}$ are
defined as (\ref{eq:overlap_CMS}) and (\ref{eq:overlap_f}) and $h_{kl}$
is given by (\ref{eq:ham_h_kl}). The vacuum of (\ref{eq:cub_vac_VW})
is the state annihilated by all lowering operators of $L$-bit and
$K$-bit systems, i.e., $F_{i}^{\left(K\right)}\ket{0}=F_{i}^{\left(L\right)}\ket{0}=0$.
In the following, we first mark remarks on the interaction vertices
in the Sec. \ref{subsec:remarks_on_vertices} and then give all the
techenical details of the derivation of (\ref{eq:cub_Vertices}) in
Sec. \ref{subsec:Derive_VW}.

\subsection{\label{subsec:remarks_on_vertices}Remarks on vertices}

The form of vertices in (\ref{eq:cub_Vertices}) can be interpreted
as follows. The prefactor $M$ of $V_{qpr}$ shows that, when a large
chain splits into two small chains of $K$ and $L$ bits, there are
$M$ ways to choose the break points, and each way contributes equally
to $V_{qpr}$. Likewise, the prefactor $KL$ of $W_{rpq}$ shows that,
when two small chains join into a large chain, there are $K\times L$
ways to choose the joint points, and each way contributes equally
to $W_{rpq}$. The operator $h_{V}^{a}=\frac{1}{2}\left(h_{K,1}^{a}+h_{M,K+1}^{a}\right)$
reflects the fact that, to break one $M$-bit string into $K$-bit
and $L$-bit strings, one needs to connect bit $1$ to bit $K$ and
bit $\left(K+1\right)$ to bit $M$. Similarly, the operator $h_{W}=h_{K,K+1}^{a}+h_{M,1}^{a}$
reflects the fact that, to join back the above two small strings into
one, one needs to connect bit $K$ to bit $\left(K+1\right)$ and
bit $M$ to bit $1$. The difference of factor $2$ between $h_{V}^{a}$
and $h_{W}^{a}$ is because that, when joining two strings, one can
inverse the labels of the first small string as $1+i\leftrightarrow K-i$
to obtain a different large string.

\subsection{\label{subsec:Derive_VW}Derivation of $V_{qpr}$ and $W_{rpq}$}

Now, let us derive the formula (\ref{eq:cub_Vertices}). Acting the
Hamiltonian to the zeroth order energy eigenstate $T_{r}\ket{0}$
and using (\ref{eq:ham_trace_energystate}) and (\ref{eq:ham_Haction_single_trace}),
we have 
\begin{align}
HT_{r}\ket{0} & =E_{r}T_{r}\ket{0}+\frac{1}{N}\int d\theta\,\sum_{i=1}^{M}\sum_{j=i+2}^{M+i}\bar{h}_{ij}T\left(\theta_{j}\cdots\theta_{i}\right)T\left(\theta_{i+1}\cdots\theta_{j-1}\right)\ket{0}\psi_{r}\left(\theta_{1},\cdots,\theta_{M}\right)\nonumber \\
 & =E_{r}T_{r}\ket{0}+\frac{1}{N}\sum_{i=1}^{M}\sum_{K=1}^{M-1}\int d\theta\,\bar{h}_{i,i+K+1}T\left(\theta_{i+K+1}\cdots\theta_{i}\right)T\left(\theta_{i+1}\cdots\theta_{i+K}\right)\ket{0}\psi_{r}\nonumber \\
 & =E_{r}T_{r}\ket{0}+\frac{1}{N}\sum_{i=1}^{M}\sum_{K=1}^{M-1}\int d\theta\,\sum_{p,q}\left(-\right)^{s\left(Kg_{p}+Lg_{q}\right)}\bar{h}_{i,i+K+1}T_{q}\bar{\psi}_{q}T_{p}\bar{\psi}_{p}\ket{0}\psi_{r},\label{eq:cub_HTr}
\end{align}
 where in the second equality we renamed the indices as $j\to i+K+1$
and in the last equality we used (\ref{eq:conj_completeness_1}).
Comparing (\ref{eq:cub_HTr}) with (\ref{eq:cub_Vertices_DefV}),
we arrive at 
\begin{equation}
\tilde{V}_{qpr}=\left(-\right)^{s\left(Kg_{q}+Lg_{p}\right)}\sum_{i=1}^{M}\int d\theta\,\bar{h}_{i,i+K+1}\bar{\psi}_{q}\left(\theta_{i+K+1}\cdots\theta_{i}\right)\bar{\psi}_{p}\left(\theta_{i+1}\cdots\theta_{i+K}\right)\psi_{r}.\label{eq:cub_Vstr}
\end{equation}
The vertex is decorated with a tilde because we have not yet applied
the constraint (\ref{eq:cub_Vertices_Sym}) to it. Note that the sign
factor is changed due to the reorder of $T_{p}$ and $T_{q}$.

The action of $H$ on the double trace produces both fusion and fission
terms: 
\begin{multline*}
HT_{p}T_{q}\ket{0}=\left(E_{p}+E_{q}\right)T_{p}T_{q}\ket{0}\\
+\frac{1}{N}\sum_{r}T_{r}\int d\theta d\eta\,\sum_{k,l}\left(-\right)^{sL\left(g_{p}-sK\right)}\bar{h}_{kl}\bar{\psi}_{r}\left(\theta_{k+1}\cdots\theta_{k}\eta_{l}\cdots\eta_{l-1}\right)\psi_{p}\left(\theta\right)\psi_{q}\left(\eta\right)\ket{0}\\
+\frac{1}{N}\sum_{r}T_{r}\int d\theta d\eta\,\sum_{k,l}\left(-\right)^{sL\left(g_{p}-sK\right)}\bar{h}_{lk}\bar{\psi}_{r}\left(\theta_{k}\cdots\theta_{k-1}\eta_{l+1}\cdots\eta_{l}\right)\psi_{p}\left(\theta\right)\psi_{q}\left(\eta\right)\ket{0}+\frac{1}{N}\text{Fission Terms}.
\end{multline*}
Comparing the above with (\ref{eq:cub_Vertices_DefW}), we have $\tilde{W}_{rpq}=W_{rpq}^{\left(1\right)}+W_{rpq}^{\left(2\right)}$,
where 
\[
W_{rpq}^{\left(1\right)}=\left(-\right)^{sL\left(g_{p}-sK\right)}\int d\theta d\eta\,\sum_{k,l}\bar{h}_{kl}\bar{\psi}_{r}\left(\theta_{k+1}\cdots\theta_{k}\eta_{l}\cdots\eta_{l-1}\right)\psi_{p}\left(\theta_{1}\cdots\theta_{K}\right)\psi_{q}\left(\eta_{1}\cdots\eta_{L}\right),
\]
\[
W_{rpq}^{\left(2\right)}=\left(-\right)^{sL\left(g_{p}-sK\right)}\int d\theta d\eta\,\sum_{k,l}\bar{h}_{lk}\bar{\psi}_{r}\left(\theta_{k}\cdots\theta_{k-1}\eta_{l+1}\cdots\eta_{l}\right)\psi_{p}\left(\theta_{1}\cdots\theta_{K}\right)\psi_{q}\left(\eta_{1}\cdots\eta_{L}\right).
\]
Note that so far the derivation of $V$ and $W$ follows the one of
ref. \cite{Thorn:2015wli} except that we changed the notation slightly
and determined the sign factors of the vertices, which are overlooked
by ref. \cite{Thorn:2015wli} in eqs. (21) and (27). 

Now, let us simplify $\tilde{V}$ and $\tilde{W}$. We denote the
integral with index $i$ in (\ref{eq:cub_Vstr}) as $\tilde{V}_{qpr}^{\left(i\right)}$.
It can be shown as follows that all the $M$ integrals $\tilde{V}_{qpr}^{\left(i\right)}$
are the same. For the integral with index $i$, we can rename all
integration variables as $\theta_{j}\to\theta_{j+1}$ and then use
the cyclic constraint (\ref{eq:ham_cyclicconstrtaint}) to bring $\psi_{r}$
and the measure to their original form. The value of the integral
is invariant under both changes but $\tilde{V}_{qpr}^{\left(i\right)}$
is changed to $\tilde{V}_{qpr}^{\left(i+1\right)}$. It implies that
$\tilde{V}_{qpr}^{\left(i\right)}$ is independent of $i$ and we
can choose $i=M$ for every integral to give 
\[
\tilde{V}_{qpr}=\left(-\right)^{s\left(Kg_{q}+Lg_{p}\right)}M\int d\theta\,\bar{\psi}_{q}\left(\theta_{K+1}\cdots\theta_{M}\right)\bar{\psi}_{p}\left(\theta_{1}\cdots\theta_{K}\right)h_{M,K+1}\psi_{r}\left(\theta_{1},\cdots,\theta_{M}\right).
\]

To find the vertex satisfying the constraint (\ref{eq:cub_Vertices_Sym}),
we let $V_{qpr}=\frac{1}{2}\left(\tilde{V}_{qpr}+\left(-\right)^{g_{p}g_{q}}\tilde{V}_{pqr}\right)$,
where $\tilde{V}_{pqr}$ can be obtained by exchanging $p\leftrightarrow q$,
$K\leftrightarrow L$:
\begin{align*}
\left(-\right)^{g_{p}g_{q}}\tilde{V}_{pqr} & =\left(-\right)^{g_{p}g_{q}+s\left(Kg_{q}+Lg_{p}\right)}M\int d\theta\,\bar{\psi}_{p}\left(\theta_{L+1}\cdots\theta_{M}\right)\bar{\psi}_{q}\left(\theta_{1}\cdots\theta_{L}\right)h_{M,L+1}\psi_{r}\\
 & =\left(-\right)^{g_{p}g_{q}+s\left(Kg_{q}+Lg_{p}\right)}M\int d\theta\,\bar{\psi}_{p}\left(\theta_{1}\cdots\theta_{K}\right)\bar{\psi}_{q}\left(\theta_{K+1}\cdots\theta_{M}\right)h_{K,1}\psi_{r}\\
 & =\left(-\right)^{s\left(Kg_{q}+Lg_{p}\right)}M\int d\theta\,\bar{\psi}_{q}\left(\theta_{K+1}\cdots\theta_{M}\right)\bar{\psi}_{p}\left(\theta_{1}\cdots\theta_{K}\right)h_{K,1}\psi_{r}.
\end{align*}
 We therefore have
\begin{equation}
V_{qpr}=\left(-\right)^{s\left(Kg_{q}+Lg_{p}\right)}M\int d\theta\,\bar{\psi}_{q}\left(\theta_{K+1}\cdots\theta_{M}\right)\bar{\psi}_{p}\left(\theta_{1}\cdots\theta_{K}\right)h_{V}\psi_{r}\left(\theta_{1},\cdots,\theta_{M}\right),\label{eq:cub_Vstr_2}
\end{equation}
 where $h_{V}$ is given by (\ref{eq:cub_hVW}).

We perform a similar calculation for the $\tilde{W}$ vertex. All
the integrals of $W^{\left(1\right)}$ and $W^{\left(2\right)}$ are
independent of the indices $k$ and $l$. So we can simply replace
the sums over $k$ and $l$ with the factor $K\times L$. We then
rename $\eta_{1},\dots,\eta_{L}$ to $\theta_{K+1},\dots,\theta_{M}$
and fix the indices as $k=K,\,l=K+1$ for $W^{\left(1\right)}$ and
$k=1,\,l=M$ for $W^{\left(2\right)}$ to give
\begin{equation}
\tilde{W}_{rpq}=\left(-\right)^{sL\left(g_{p}-sK\right)}KL\int d\theta\,\bar{\psi}_{r}\left(\theta_{1}\cdots\theta_{M}\right)\left(h_{K,K+1}+h_{M,1}\right)\psi_{p}\left(\theta_{1}\cdots\theta_{K}\right)\psi_{q}\left(\theta_{K+1}\cdots\theta_{M}\right).\label{eq:cub_Wrst}
\end{equation}
Exchanging $p\leftrightarrow q$ and $K\leftrightarrow L$, we have
\begin{equation}
\tilde{W}_{rqp}=\left(-\right)^{sK\left(g_{q}-sL\right)}KL\int d\theta\,\bar{\psi}_{r}\left(\theta_{1}\cdots\theta_{M}\right)\left(h_{L,L+1}+h_{M,1}\right)\psi_{q}\left(\theta_{1}\cdots\theta_{L}\right)\psi_{p}\left(\theta_{L+1}\cdots\theta_{M}\right).
\end{equation}
Renaming the integral variables as $\left\{ \theta_{1},\cdots,\theta_{L}\right\} \to\left\{ \theta_{K+1},\cdots,\theta_{M}\right\} $,
$\left\{ \theta_{L+1},\cdots,\theta_{M}\right\} \to\left\{ \theta_{1},\cdots,\theta_{K}\right\} $,
under which $h_{L,L+1}+h_{M,1}$ becomes $h_{M,1}+h_{K,K+1}$, and
then applying the property that $\bar{\psi}_{r}\left(\theta_{1}\cdots\theta_{M}\right)$
is invariant under the cyclic permutation $\theta_{k}\to\theta_{k+1}$\footnote{One can show that $\bar{\psi}_{r}\left(\theta_{1}\cdots\theta_{M}\right)$
is invariant under the cyclic permutation $\theta_{k}\to\theta_{k+1}$
as follows. From eq. (\ref{eq:conj_psibarG:1}) and (\ref{eq:conj_psibarG:2}),
we see that $\bar{\psi}_{G}^{s=1}\to\left(-\right)^{M-1}\bar{\psi}_{G}^{s=1}$
as $\theta_{k}\to\theta_{k+1}$. It then follows that $\bar{\psi}_{G}$
transforms as $\bar{\psi}_{G}\to\left(-\right)^{s\left(M-1\right)}\bar{\psi}_{G}$.
From the cyclic constraint (\ref{eq:dia_cyclic_constraint_s}), we
see that $\tilde{F}_{\left\{ r\right\} }$ transforms in the same
way as $\bar{\psi}_{G}$. Therefore, $\bar{\psi}_{r}=\tilde{F}_{\left\{ r\right\} }\bar{\psi}_{G}$
is invariant.}, we obtain that $\tilde{W}_{rqp}=\left(-\right)^{g_{p}g_{q}}\tilde{W}_{rpq}$,
which implies that $W_{rpq}=\left(\tilde{W}_{rpq}+\left(-\right)^{g_{p}g_{q}}\tilde{W}_{rqp}\right)=\tilde{W}_{rpq}$.

Let us now get rid of the integral in the expression of $V$. For
simplicity, we consider the $s=1$ case. We use (\ref{eq:dia_general_eigenfunction_s})
and (\ref{eq:conj_psibar_general}) to write $\psi_{r}=F_{\left\{ r\right\} }^{\dagger}\psi_{G}^{\left(M\right)}$,
$\bar{\psi}_{r}=\tilde{F}_{\left\{ r\right\} }\bar{\psi}_{G}^{\left(M\right)}$
and similarly for states $p$ and $q$. We then use (\ref{eq:overlap_psi_M1})
to express $\psi_{G}^{\left(M\right)}$ in terms of $\psi_{G}^{\left(L\right)}$
and $\psi_{G}^{\left(K\right)}$. By a little algebra, we arrive at
\begin{equation}
V_{qpr}^{s=1}=\left(-\right)^{L\left(g_{p}-K\right)}M\left|\det C\right|^{1/2}\int d\theta\,\bar{\psi}_{G}^{\left(L\right)}\bar{\psi}_{G}^{\left(K\right)}F_{\left\{ qp\right\} }h_{V}F_{\left\{ r\right\} }^{\dagger}\exp\left(\frac{1}{2}\sum_{kl}f_{k}^{\dagger}D_{kl}f_{l}^{\dagger}\right)\psi_{G}^{\left(K\right)}\psi_{G}^{\left(L\right)}.\label{eq:cub_Vqpr_3}
\end{equation}
The ground eigenfunctions $\psi_{G}^{\left(L\right)}$ and $\psi_{G}^{\left(K\right)}$
are annihilated by any lowering eigenoperators of the small chains.
Their conjugates $\bar{\psi}_{G}^{\left(L\right)}$ and $\bar{\psi}_{G}^{\left(K\right)}$
can be annihilated by any raising eigenoperators of the small chains,
as eq. (\ref{eq:App_psiG_FtildeG}) shows. Therefore, the rhs of (\ref{eq:cub_Vqpr_3})
can be interpreted as a vacuum expectation value of the operator $F_{\left\{ q\right\} }F_{\left\{ p\right\} }h_{V}F_{\left\{ r\right\} }^{\dagger}\exp\left(\frac{1}{2}\sum_{kl}f_{k}^{\dagger}D_{kl}f_{l}^{\dagger}\right)$.
We therefore have 
\[
V_{qpr}^{s=1}=\left(-\right)^{L\left(g_{p}-K\right)}M\left|\det C\right|^{1/2}\left\langle F_{\left\{ qp\right\} }h_{V}F_{\left\{ r\right\} }^{\dagger}\exp\left(\frac{1}{2}\sum_{kl}f_{k}^{\dagger}D_{kl}f_{l}^{\dagger}\right)\right\rangle ,
\]
 where the vacuum is understood to be the state annihilated by all
$F_{i}^{\left(K\right)}$ and $F_{i}^{\left(L\right)}$. We perform
a similar calculation for $W_{rpq}$ and find 
\begin{align*}
W_{rpq}^{s=1} & =\left(-\right)^{L\left(g_{p}-K\right)}KL\left|\det C\right|^{\frac{1}{2}}\left\langle \exp\left(\frac{1}{2}f_{k}D_{kl}^{\dagger}f_{l}\right)F_{\left\{ r\right\} }h_{W}F_{\left\{ p\right\} }^{\dagger}F_{\left\{ q\right\} }^{\dagger}\right\rangle \\
 & =\left(-\right)^{L\left(g_{p}-K\right)}KL\left|\det C\right|^{\frac{1}{2}}\left\langle F_{\left\{ qp\right\} }h_{W}F_{\left\{ r\right\} }^{\dagger}\exp\left(\frac{1}{2}f_{k}^{\dagger}D_{kl}f_{l}^{\dagger}\right)\right\rangle ^{\dagger}.
\end{align*}

Note that $V_{qpr}^{s=1}$ and $W_{rpq}^{s=1}$ have the same sign
factor $\left(-\right)^{L\left(g_{p}-K\right)}$. We shall see that
physical observables, like one-loop self-energies, only depend on
products like $W_{rpq}V_{qpr}$. It implies that the sign factors
are unphysical and can be dropped in the calculation of physical observables.
So, for arbitrary $s$, up to a common unphysical sign factor, we
can express $V$ and $W$ as products of vacuum expectation values
over spin index $a$. We therefore obtain the formula (\ref{eq:cub_Vertices}).

To calculate the vacuum expectation values, we need to express $h_{V}$
and $h_{W}$ in terms of eigenoperators. From eqs. (\ref{eq:App_hkl_h}),
(\ref{eq:App_hkl_Anm}), and (\ref{eq:App_hkl_mu}), we have 
\begin{eqnarray}
h_{\left(V,W\right)} & = & \frac{2}{M}\sum_{n,m=0}^{M-1}\left(A_{nm}^{\left(V,W\right)\dagger}F_{n}^{\dagger}F_{m}^{\dagger}+A_{nm}^{\left(V,W\right)}F_{n}F_{m}+2A_{-n,m}^{\left(V,W\right)}F_{n}^{\dagger}F_{m}\right)+\frac{2}{M}\mu_{\left(V,W\right)},\label{eq:cub_h_kl}
\end{eqnarray}
 where\begin{subequations}\label{eq:cub_Amu} 
\begin{align}
A_{nm}^{\left(V\right)} & =\frac{1}{2}\left[1-\exp\left(2\pi i\frac{Kn}{M}\right)\right]\left[1-\exp\left(2\pi i\frac{Km}{M}\right)\right]\sin\frac{m-n}{2M}\pi\nonumber \\
 & +\frac{1}{2}\left[\exp\left(2\pi i\frac{Kn}{M}\right)+\exp\left(2\pi i\frac{Km}{M}\right)\right]\left[1+\exp\left(\pi i\frac{m+n}{M}\right)\right]\sin\frac{m-n}{2M}\pi\\
\mu_{V} & =-\cot\frac{\pi}{2M}+\frac{1}{2}\left(\cot\frac{2K-1}{2M}\pi-\cot\frac{2K+1}{2M}\pi\right)+M\xi,\\
A_{nm}^{\left(W\right)} & =\left[1+\exp\left(\pi i\frac{n+m}{M}\right)\right]\left[1+\exp\left(2\pi iK\frac{m+n}{M}\right)\right]\sin\frac{m-n}{2M}\pi,\\
\mu_{W} & =-4\cot\frac{\pi}{2M}+2M\xi.
\end{align}
\end{subequations}

\section{\label{sec:EnergyCorrection}One-loop self-energy}

One application of the interaction vertices is to calculate the one-loop
self-energy, i.e., the $\mathcal{O}\left(1/N^{2}\right)$ correction
to energy spectrum. In this section, we will first express the one-loop
self-energy in terms of cubic interaction vertices \cite{Thorn:2015wli}.
We then apply the results of previous sections and obtain a formula
for analytic and numerical computation.

For a finite $N$ energy eigenstate, we use the ansatz 
\begin{equation}
\ket{E}=T_{r}\ket{0}+T_{p}T_{q}\ket{0}C_{pq}+\cdots,
\end{equation}
 where the coefficients $C_{pq}=\left(-\right)^{g_{p}g_{q}}C_{qp}$
are c-numbers of order $1/N$. Imposing the eigenvalue equation $\left(H-E\right)\ket{E}=0$
and using perturbation theory, we obtain \cite{Thorn:2015wli}
\begin{align}
C_{pq} & =\frac{1}{E_{r}-E_{p}-E_{q}}\frac{1}{N}V_{qpr}+\mathcal{O}\left(N^{-2}\right),\\
\Delta E_{r} & =\frac{1}{N^{2}}\sum_{K=1}^{M-1}\sum_{p,q}W_{rpq}\frac{1}{E_{r}-E_{p}-E_{q}}V_{qpr},\label{eq:olp_deltaE0}
\end{align}
 where $\Delta E_{r}$ is the leading order correction to $E_{r}$,
i.e., $E=E_{r}+\Delta E_{r}+\mathcal{O}\left(1/N^{3}\right)$. We
stress that the vertices in (\ref{eq:olp_deltaE0}) should be the
ones satisfying the constraint (\ref{eq:cub_Vertices_Sym}); otherwise,
it would lead to an incorrect $\Delta E_{r}$.

We now apply the formulas of $V$ and $W$ to (\ref{eq:olp_deltaE0}).
Let us first consider the $s=1$ case. The zero modes require special
treatment. Substitute (\ref{eq:cub_VertexV}) and (\ref{eq:cub_VertexW})
into (\ref{eq:olp_deltaE0}) and write the sum over zero modes explicitly,
\[
\Delta E_{r}^{s=1}=\frac{1}{N^{2}}\sum_{K=1}^{M-1}{\sum_{p,q}}^{\prime}\frac{KLM\left|\det C\right|}{E_{r}-E_{p}-E_{q}}\sum_{\lambda,\kappa=0,1}\left\langle F_{\left\{ qp\right\} }F_{0,K}^{\lambda}F_{0,L}^{\kappa},F_{\left\{ r\right\} }^{\dagger}\right\rangle _{W}^{*}\left\langle F_{\left\{ qp\right\} }F_{0,K}^{\lambda}F_{0,L}^{\kappa},F_{\left\{ r\right\} }^{\dagger}\right\rangle _{V}
\]
 where we wrote $F_{0}^{\left(K\right)}$ as $F_{0,K}$ for convenience
and ${\sum_{p,q}}^{\prime}$ indicates the sum over states without
zero modes. We can replace $F_{0}^{\left(K\right)}$ and $F_{0}^{\left(L\right)}$
with $f_{0}$ and $f_{M-1}$ given the following reasoning. The sum
over $\lambda$ and $\kappa$ produces four terms. For the term with
$\lambda=\kappa=1$, we find $F_{0}^{\left(K\right)}F_{0}^{\left(L\right)}=e^{i\pi/4}f_{0}f_{M-1}$
by eqs. (\ref{eq:overlap_f:1}) and (\ref{eq:overlap_f:4}). The phase
is irrelevant. The $\lambda=1,\,\kappa=0$ and $\lambda=0,\,\kappa=1$
terms are quadratic forms of $F_{0}^{\left(K\right)}$ and $F_{0}^{\left(L\right)}$.
One can easily verify that $F_{0}^{\left(K\right)*}F_{0}^{\left(K\right)}+F_{0}^{\left(L\right)*}F_{0}^{\left(L\right)}=f_{0}^{*}f_{0}+f_{M-1}^{*}f_{M-1}$.
So, the sum over $F_{0}^{\left(K\right)}$ and $F_{0}^{\left(L\right)}$
 can be replaced by the one over $f_{0}$ and $f_{M-1}$. We then
have 
\[
\Delta E_{r}^{s=1}=\frac{1}{N^{2}}\sum_{K=1}^{M-1}{\sum_{p,q}}^{\prime}\frac{KLM\left|\det C\right|}{E_{r}-E_{p}-E_{q}}\sum_{\lambda_{i}=0,1}\left\langle F_{\left\{ qp\right\} }f_{0}^{\lambda_{0}}f_{M-1}^{\lambda_{M-1}},F_{\left\{ r\right\} }^{\dagger}\right\rangle _{W}^{*}\left\langle F_{\left\{ qp\right\} }f_{0}^{\lambda_{0}}f_{M-1}^{\lambda_{M-1}},F_{\left\{ r\right\} }^{\dagger}\right\rangle _{V}.
\]
For arbitrary $s$, $\left|\det C\right|$ is replaced by $\left|\det C\right|^{s}$,
and each term inside the summation becomes a product over $a$. So,
we have 
\[
\Delta E_{r}=\frac{1}{N^{2}}\sum_{K=1}^{M-1}{\sum_{p,q}}^{\prime}\frac{KLM\left|\det C\right|^{s}}{E_{r}-E_{p}-E_{q}}\sum_{\lambda_{i,j}=0,1}\prod_{a=1}^{s}\left\langle F_{\left\{ qp\right\} }^{a}f_{0,a}^{\lambda_{a,0}}f_{M-1,a}^{\lambda_{a,M-1}},F_{\left\{ r\right\} }^{a\dagger}\right\rangle _{W}^{*}\left\langle F_{\left\{ qp\right\} }^{a}f_{0,a}^{\lambda_{a,0}}f_{M-1,a}^{\lambda_{a,M-1}},F_{\left\{ r\right\} }^{a\dagger}\right\rangle _{V}.
\]
Note that the sum over $\lambda_{i,j}$ can be performed for each
$a$ independently. So, we can move the sum over $\lambda_{i,j}$
inside the product over $a$ to give
\begin{equation}
\Delta E_{r}=\frac{1}{N^{2}}\sum_{K=1}^{M-1}{\sum_{p,q}}^{\prime}\frac{KLM\left|\det C\right|^{s}}{E_{r}-E_{p}-E_{q}}\prod_{a=1}^{s}\left(\sum_{i=1}^{4}\left\langle F_{\left\{ qp\right\} }^{a}Z_{i}^{a},F_{\left\{ r\right\} }^{a\dagger}\right\rangle _{W}^{*}\left\langle F_{\left\{ qp\right\} }^{a}Z_{i}^{a},F_{\left\{ r\right\} }^{a\dagger}\right\rangle _{V}\right),\label{eq:olp_DeltaE}
\end{equation}
 where $Z^{a}=\left(1,\,f_{0}^{a},\,f_{M-1}^{a},\,f_{0}^{a}f_{M-1}^{a}\right)$. 

\subsection{Ground energy correction}

In principle, we can now calculate one-loop self-energy for any single
trace energy state with eq. (\ref{eq:olp_DeltaE}). But in general,
the calculation is tedious. Let us consider the simplest case that
$\psi_{r}$ is the ground state, i.e., $F_{\left\{ r\right\} }=1$.
For convenience, we denote $\left\langle O,1\right\rangle _{V,W}$
as $\left\langle O\right\rangle _{V,W}$. We only consider the $s=1$
case here, since $s>1$ cases are simply products of the $s=1$ case. 

We need to calculate the vacuum expectation value $\left\langle \cdots h\exp\left(\frac{1}{2}f_{k}^{\dagger}D_{kl}f_{l}^{\dagger}\right)\right\rangle $.
In terms of eigenoperators, $h_{kl}$ contains quadratic terms of
the form $A_{nm}^{\dagger}F_{m}^{\dagger}F_{n}^{\dagger}$, $A_{-n,m}F_{m}^{\dagger}F_{n}$,
and $A_{nm}F_{n}F_{m}$ and a constant term $\mu$, as eq. (\ref{eq:cub_h_kl})
shows. Since $F_{m}\exp\left(\frac{1}{2}f_{k}^{\dagger}D_{kl}f_{l}^{\dagger}\right)\psi_{G}^{\left(K\right)}\psi_{G}^{\left(L\right)}=F_{m}\psi_{G}^{\left(M\right)}=0$,
only the $F^{\dagger}F^{\dagger}$ and the constant terms make a nonzero
contribution: 
\[
\left\langle \cdots h\exp\left(\frac{1}{2}f_{k}^{\dagger}D_{kl}f_{l}^{\dagger}\right)\right\rangle =\frac{2}{M}\left\langle \cdots\left(A_{nm}^{\dagger}F_{n}^{\dagger}F_{m}^{\dagger}+\mu\right)\exp\left(\frac{1}{2}f_{k}^{\dagger}D_{kl}f_{l}^{\dagger}\right)\right\rangle .
\]
To calculate the result of $A_{nm}^{\dagger}F_{n}^{\dagger}F_{m}^{\dagger}$
term, we need to express $F_{m}$ in terms of a linear combination
of $f_{k}$ and $f_{k}^{\dagger}$, as (\ref{eq:overlap_Ff_relation})
shows, and commute $f_{k}$ through the exponential. This is done
in Appendix \ref{sec:app_AFF}. Using eq. (\ref{eq:App_AFF_1}), we
have
\begin{equation}
\left\langle F_{\left\{ qp\right\} }Z_{i}\right\rangle _{V,W}=\frac{2}{M}\left(\mu_{V,W}^{\prime}+B_{mn}^{\left(V,W\right)}\frac{\partial}{\partial D_{mn}}\right)\left\langle F_{\left\{ qp\right\} }Z_{i}\exp\left(\frac{1}{2}f_{k}^{\dagger}D_{kl}f_{l}^{\dagger}\right)\right\rangle ,\label{eq:olp_vev_v}
\end{equation}
 where 
\[
\mu_{V,W}^{\prime}=\mu_{V,W}-\tr\left(S^{*}C^{-1}A_{V,W}^{\dagger}\right),\quad B_{V,W}=C^{-1}A_{V,W}^{\dagger}\left(C^{-1}\right)^{T}
\]
 with $\mu_{V,W}$ and $A_{V,W}$ defined in (\ref{eq:cub_Amu}).
Finally, the vacuum expectation values on the rhs of (\ref{eq:olp_vev_v})
can be calculated using
\[
\left\langle f_{i_{1}}f_{i_{2}}\cdots f_{i_{2n-1}}f_{i_{2n}}\exp\left(\frac{1}{2}f_{k}^{\dagger}D_{kl}f_{l}^{\dagger}\right)\right\rangle =\left(-\right)^{n}{\sum_{P\in S_{2n}}}^{\prime}\left(-\right)^{P}D_{i_{P(1)}i_{P(2)}}D_{i_{P(3)}i_{P(4)}}\cdots D_{i_{P(2n-1)}i_{P(2n)}},
\]
 where $S_{2n}$ is the set of all permutations of $2n$ integers,
$\left(-\right)^{P}$ is the signature of permutation $P$, and ${\sum}^{\prime}$
indicates the sum over permutations satisfying 
\begin{align*}
P\left(1\right) & <P\left(2\right),\quad P\left(3\right)<P\left(4\right),\cdots,P\left(2n-1\right)<P\left(2n\right),\\
P\left(1\right) & <P\left(3\right)<P\left(5\right)<\cdots<P\left(2n-1\right).
\end{align*}

Combining the above together, we can calculate the one-loop self-energy
of the ground state. As the complete formula is very complicated,
we do not bother writing it here. In Appendix \ref{sec:app_example},
we show examples of using formula (\ref{eq:olp_DeltaE}) to calculate
the one-loop self-energies of the $M=3,\,s=1$ and $M=3,\,s=2$ cases.
For $M=3,\,s=1$, we have
\[
\Delta E_{G}=\frac{1}{N^{2}}\left[-3\left(3\sqrt{3}-5\right)\xi^{2}+2\left(12-7\sqrt{3}\right)\xi-\frac{3}{2}\left(3\sqrt{3}-5\right)\right],
\]
 and for $M=3,\,s=2$, we have 
\[
\Delta E_{G}=\frac{1}{N^{2}}\left(-66\sqrt{3}\xi^{4}+360\xi^{3}-230\sqrt{3}\xi^{2}+180\xi-\frac{33\sqrt{3}}{2}\right).
\]
In general, $\Delta E_{G}$ is a polynomial of $\xi$ of degree $2s$.

\subsection{Large $M$ behavior}

We conclude this section by considering the large $M$ behavior of
$\Delta E_{G}$\footnote{The large $M$ discussion is mainly based on comments by Charles Thorn.}.
The vacuum expectation values in (\ref{eq:olp_DeltaE}) only depends
on the ratio $K/M$ and therefore can be considered as $\mathcal{O}\left(1\right)$.
So, when $M$ is large, 
\begin{equation}
\Delta E_{G}\sim\frac{1}{N^{2}}\sum_{K=1}^{M-1}{\sum_{p,q}}^{\prime}\frac{KLM\left|\det C\right|^{s}}{E_{G}-E_{p}-E_{q}}.\label{eq:olp_largeM}
\end{equation}
In (\ref{eq:olp_largeM}), the factor $KLM$ scales as $M^{3}$, $\left|\det C\right|$
scales as $M^{-s/4}$ by eq. (\ref{eq:overlap_detC}), and the sum
over $K$ gives another factor of $M$. These three parts produce
a factor scale as $M^{4-s/4}$. 

We then consider the large $M$ behavior of $1/\left(E_{G}-E_{p}-E_{q}\right)$.
When $s$ is even, both $p$ and $q$ can be ground states, and hence
$1/\left(E_{G}-E_{p}-E_{q}\right)\sim\mathcal{O}\left(M\right)$ by
eq. (\ref{eq:diag_ground_energy}). When $s$ is odd, $M$ has to
be odd in order to have the physical $M$-bit ground state, and one
of the small strings must have an even bit number. It implies that
the ground state of one small chain is forbidden by the cyclic constraint
(\ref{eq:dia_cyclic_constraint_s}). Therefore, $1/\left(E_{G}-E_{p}-E_{q}\right)\sim\mathcal{O}\left(1\right)$
for odd $s$. 

Combining the above together, we have
\begin{equation}
\Delta E_{G}\sim\begin{cases}
M^{5-s/4} & \text{for even }s\\
M^{4-s/4} & \text{for odd }s
\end{cases}.\label{eq:largeM_DeltaE}
\end{equation}
In analogy with the standard string theory, we can infer from eq.
(\ref{eq:largeM_DeltaE}) the critical Grassmann dimension of the
model, where Lorentz invariance in $1+1$ dimensions is regained.
In the lightcone coordinates, $P^{+}$ is identified as $mM$, and
$P^{-}$ is identified as $E$. So, the Poincar invariant dispersion
relation $P^{-}\sim1/P^{+}$ implies $E\sim1/M$. Therefore, the Lorentz
invariance requires $s=24$. The model in the special $s=24$ case
is called the protostring model\cite{Thorn:2015wli}.

\section{\label{sec:Numerical-Results}Numerical results}

We have derived a formula for the one-loop correction to the ground
energy. As Appendix \ref{sec:app_example} shows, however, the calculation
is tedious even for the simplest case. We therefore turn to numerical
computation\footnote{The source code for the numerical computation can be found in ref.
\cite{Chen:2015GitHub}.}. As the complexity of the calculation grows dramatically, the highest
$M$ for which we performed numerical computation is 27 for $s=1$
and 16 for $s=2$ and continues decreasing as $s$ increases. Since
only the ground energy is considered, we will simply write the ground
energy as $E$ and its correction as $\Delta E$ and also suppress
the $1/N^{2}$ factor.

\begin{figure}
\begin{centering}
\includegraphics[width=0.5\textwidth]{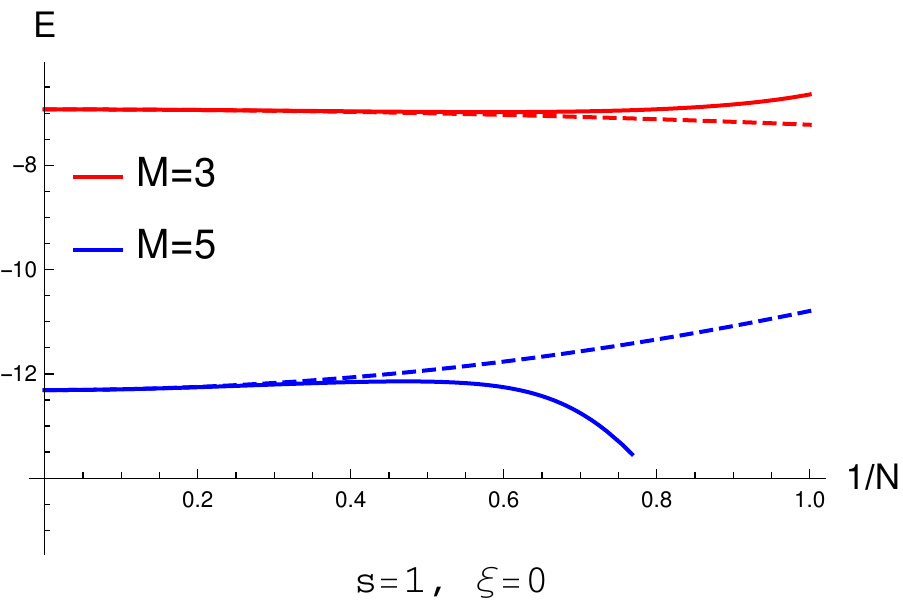}\includegraphics[width=0.5\textwidth]{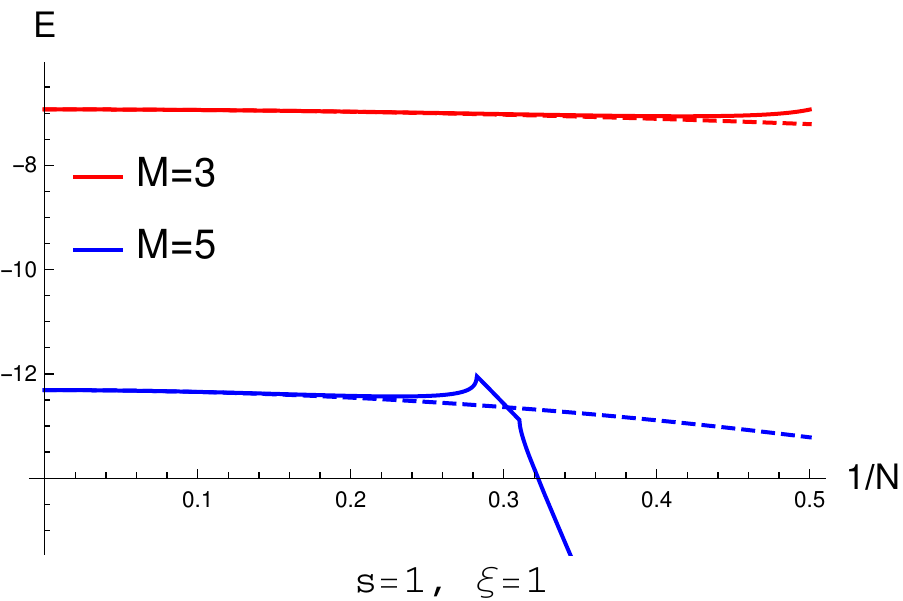}
\par\end{centering}
\caption{\label{fig:E_compare}Ground energy as a function of $1/N$ for the
$s=1,\,\xi=0$ and $s=1.\,\xi=1$ cases. The solid lines plot the
exact numerical results by the method of ref. \cite{Chen:2016hkz}.
The dashed lines plot the $1/N^{2}$ order perturbation results.}
\end{figure}

We first compare the perturbation results with the exact numerical
results, which are obtained by the method of ref. \cite{Chen:2016hkz}.
Figure \ref{fig:E_compare} plots the change of ground energy with
respect to the $1/N$ for $M=3$ and $5$ in the $s=1$ case. The
solid lines are exact numerical results, and the dashed lines are
$\mathcal{O}\left(1/N^{2}\right)$ perturbation results. We see that
the two types of results match very well for $N$ large enough. One
interesting observation is that, when $N$ is small, the perturbation
results of $M=3$ are lower than the exact results, while the perturbation
results of $M=5$ are above the exact results. It implies that the
$\mathcal{O}\left(1/N^{4}\right)$ correction is positive for $M=3$
and negative for $M=5$. 

We then verify the large $M$ behavior of $\Delta E$. Instead of
plotting $\Delta E$ with respect to $M$, we study its ``inner structure'',
that is the contribution of each $K$ to $\Delta E$, denoted by $\Delta E_{i}$
and defined as 
\[
\Delta E=\sum_{K=1}^{M-1}\Delta E_{i},\quad i=\frac{K}{M}.
\]
Since the power of $M$ in the large $M$ behavior of $\Delta E_{i}$
is 1 lower than that of $\Delta E$, we introduce the normalized $\Delta E_{i}$
to remove the $M$ dependence: 
\[
\Delta\hat{E}_{i}=\begin{cases}
\Delta E_{i}M^{-4+s/4} & \quad\text{for even }s\\
\Delta E_{i}M^{-3+s/4} & \quad\text{for odd }s
\end{cases}.
\]
We expect that, for fixed $s$ and $\xi$, $\Delta\hat{E}_{i}$ only
depends on the ratio $K/M$. 

\begin{figure}
\begin{centering}
\includegraphics[width=0.5\textwidth]{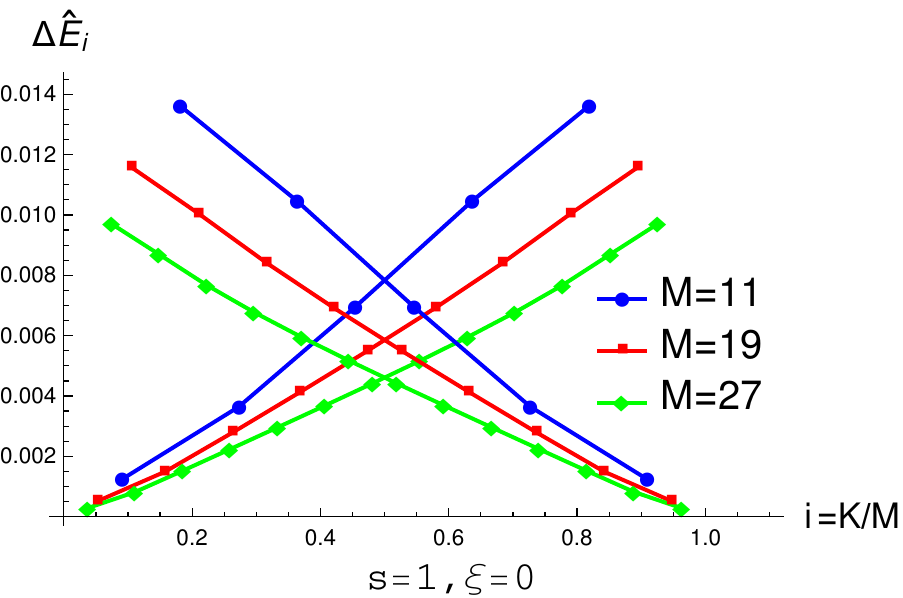}\includegraphics[width=0.5\textwidth]{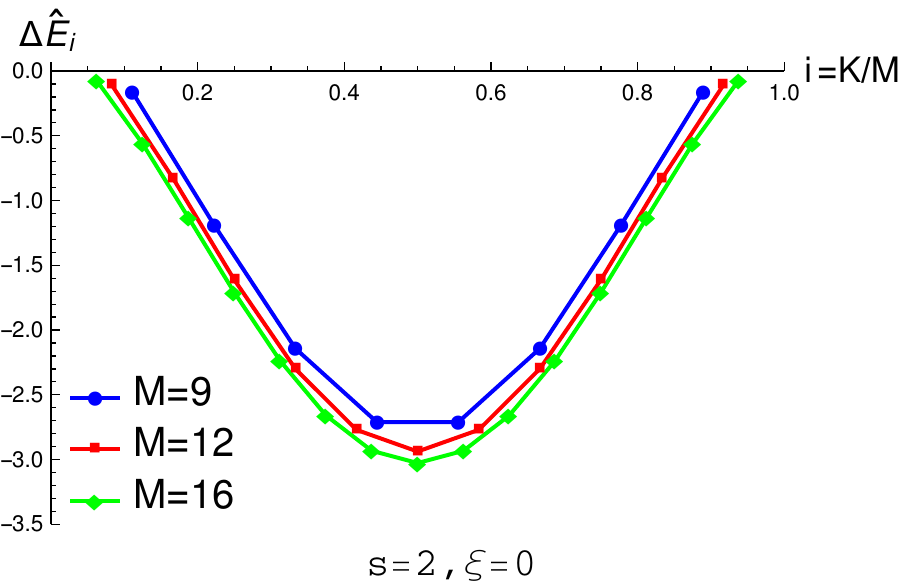}
\par\end{centering}
\begin{centering}
\includegraphics[width=0.5\textwidth]{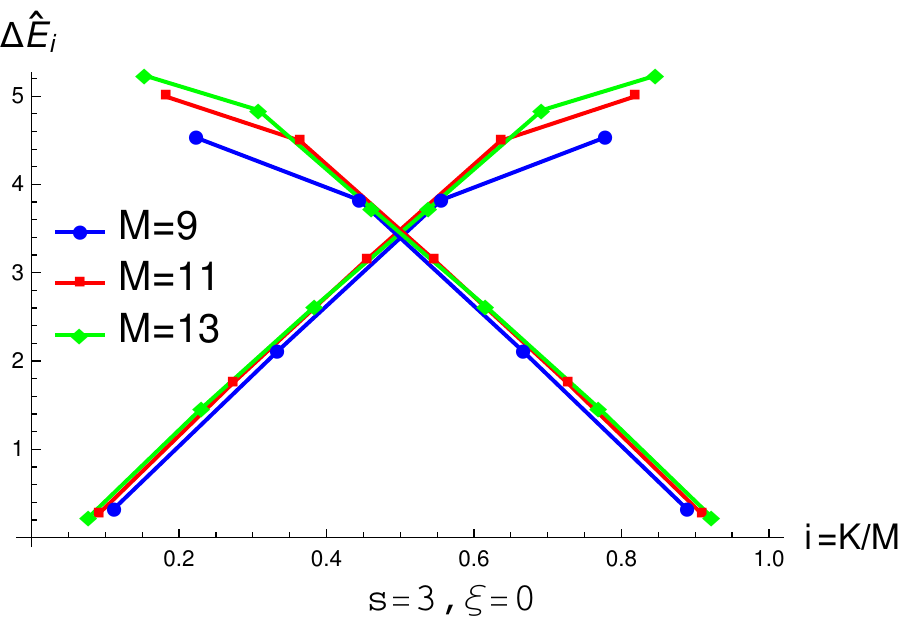}\includegraphics[width=0.5\textwidth]{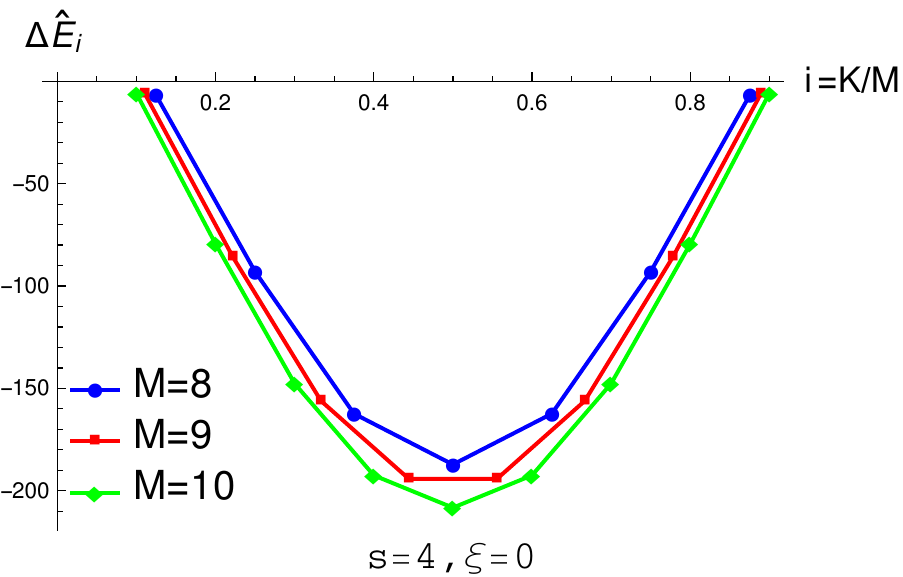}
\par\end{centering}
\caption{\label{fig:SubEvsMxi0}$\Delta\hat{E}_{i}$ as a function of $i=K/M$.
For odd $s$ cases, the curves above the horizontal axis are for odd
$K$ points, and the curves below are for even $K$ points.}
\end{figure}

The plots of $\Delta\hat{E}_{i}$ as a function of $i=K/M$ are shown
in Fig. \ref{fig:SubEvsMxi0}, where $\xi=0$ for all four plots.
When $s$ is odd, only odd values of $M$ are allowed and each $M$
has two curves, one for odd $K$ points and the other one for even
$K$ points, for a reason will be clear shortly. For $s=2,\,3,\,4$
cases, the curves of different $M$ values are very close to each
other, so the asymptotic behavior is evident. For the $s=1$ case,
the gaps between consecutive curves become smaller as $M$ increases,
which is consistent with the expected asymptotic behavior. It is therefore
fair to conclude that the large $M$ behavior is confirmed.

The fact that there are two curves for each $M$ in odd $s$ cases
can be understood as follows. Let us consider the $s=1$ case and
take examples of $K=1$ and $K=2$, where the former has a much lower
contribution to $\Delta E$ than the latter according to the plots.
Assuming that $M$ is large enough, we have the other small chains
with bit number $L\gg K$. Since $M$ is odd, $L$ is even for $K=1$
and odd for $K=2$. The lowest energies of these two cases, which
are equal to $-4\cot\frac{\pi}{2L}-4\cos\frac{\pi}{2K}+8$ according
to (\ref{eq:dia_general_energy_s}) and the cyclic constraint (\ref{eq:dia_cyclic_constraint_s}),
differ only by $\mathcal{O}\left(1\right)$. Now, we compare these
two cases in the low energy regime, in which the gap between energy
levels and the lowest energies are at most of order $1/M$. Consider
the numbers of states in the low energy regime. Because of the cyclic
constraint, only chains with an even bit number have excited states
with energy gaps of order $1/M$ above the lowest energy. For $K=1$,
the number of states in the low energy regime roughly equals $P\left(L/2\right)$,
the partition number of $L/2$; for $K=2$, it equals $P\left(2/2\right)=1$.
It implies that the low energy regime of $K=1$ is much denser than
the one of $K=2$. Therefore, for large enough $M$, the $K=1$ case
has much lower average energy than the $K=2$ case. This reasoning
holds when $K$ is small. Hence, small odd $K$ cases have a lower
contribution to $\Delta E$ than small even $K$. 

\begin{figure}
\begin{centering}
\includegraphics[width=0.5\textwidth]{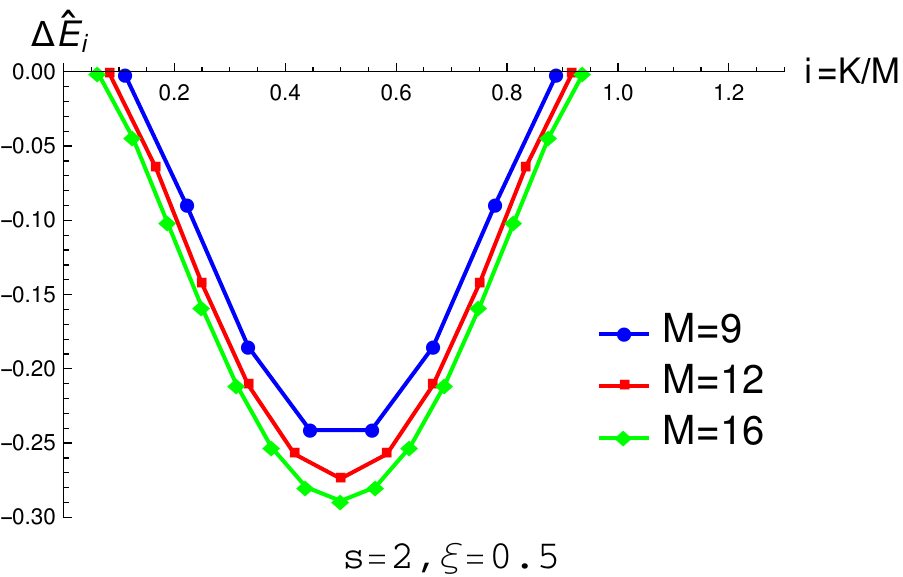}\includegraphics[width=0.5\textwidth]{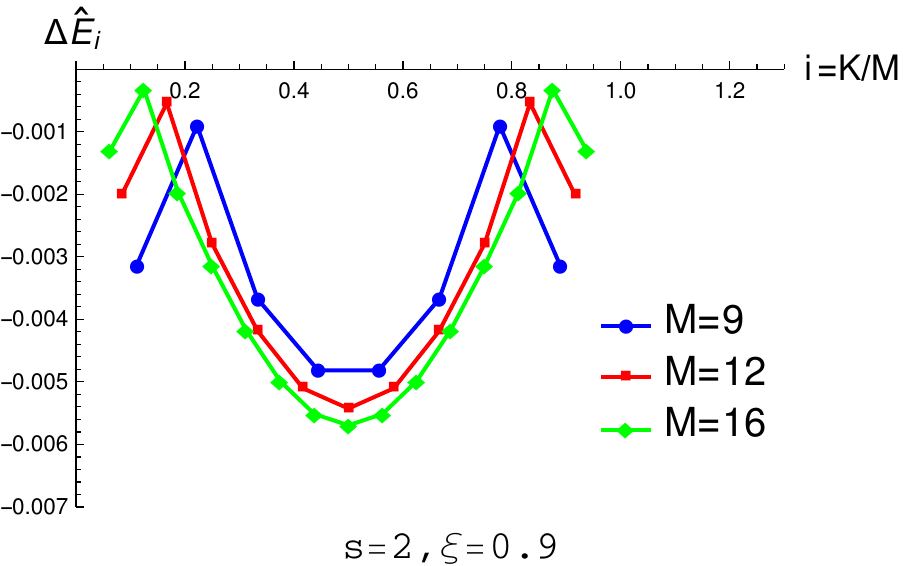}
\par\end{centering}
\begin{centering}
\includegraphics[width=0.5\textwidth]{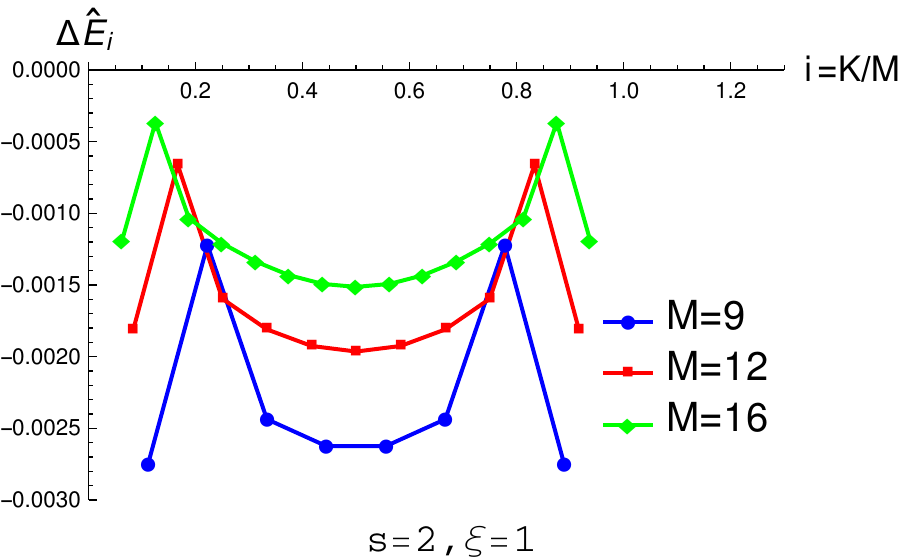}\includegraphics[width=0.5\textwidth]{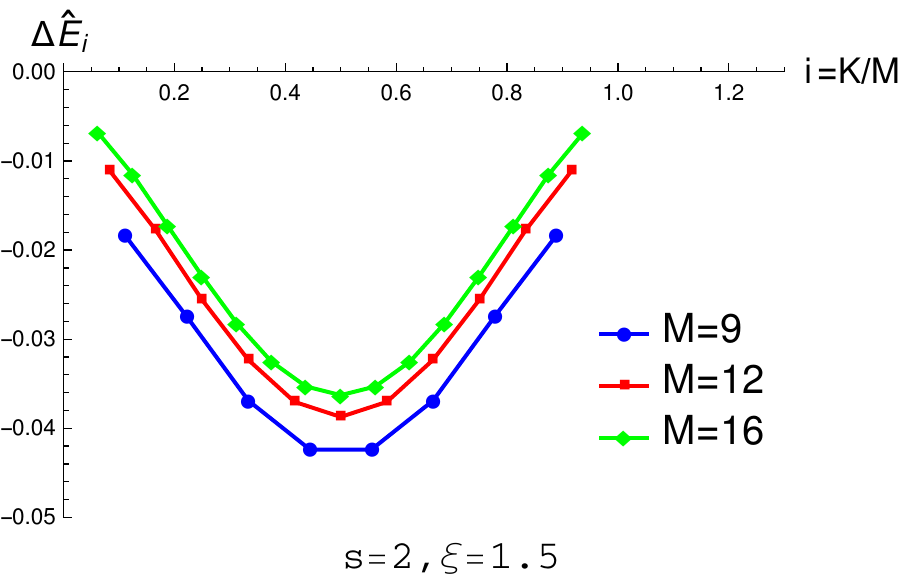}
\par\end{centering}
\caption{\label{fig:SubEvsMs2}$\Delta\hat{E}_{i}$ as a function of $i=K/M$
for the cases of $s=2$.}
\end{figure}

We next consider the effect of the $\xi$ parameter. Figure \ref{fig:SubEvsMs2}
shows the plots of $\Delta\hat{E}_{i}$ with respect to $i=K/M$ for
$s=2$ with different values of $\xi$. From the plots, the $\xi=0.5$
and $\xi=1.5$ cases show a smooth asymptotic behavior as the cases
in Fig. \ref{fig:SubEvsMxi0}. But when $\xi$ is close to $1$, curves
are not smooth and intersect each other. When $\xi<1$, the curve
moves downward as $M$ increases, which implies that $\Delta E$ decreases
as $M$ increases. So, $\Delta E$ is not bounded from below, and
the system is not stable. In contrast, when $\xi>1$, the curve moves
upward as $M$ increases, which implies a stable system. This is related
to a special feature of the $\xi=1$ case. Recall that the Hamiltonian
has an $H_{1}$ part shown as (\ref{eq:App_ham_subh:1}). This part
produces a term like $-s\tr\bar{\phi}_{12\cdots s}\bar{\phi}_{12\cdots s}\phi_{12\cdots s}\phi_{12\cdots s}$.
When $s$ is even, $\phi_{12\cdots s}$ is a scalar and this term
behaves like a scalar potential with a negative coefficient, which
leads to a dangerous instability. But when $\xi=1$, this term is
canceled exactly by $s\xi\Delta H$. That being said, for even $s$,
$\xi=1$ is the minimal value for the potential to be bounded from
below. To build a physical string bit model for even $s$, we should
require $\xi\geq1$. 

\begin{figure}
\begin{centering}
\includegraphics[width=0.5\textwidth]{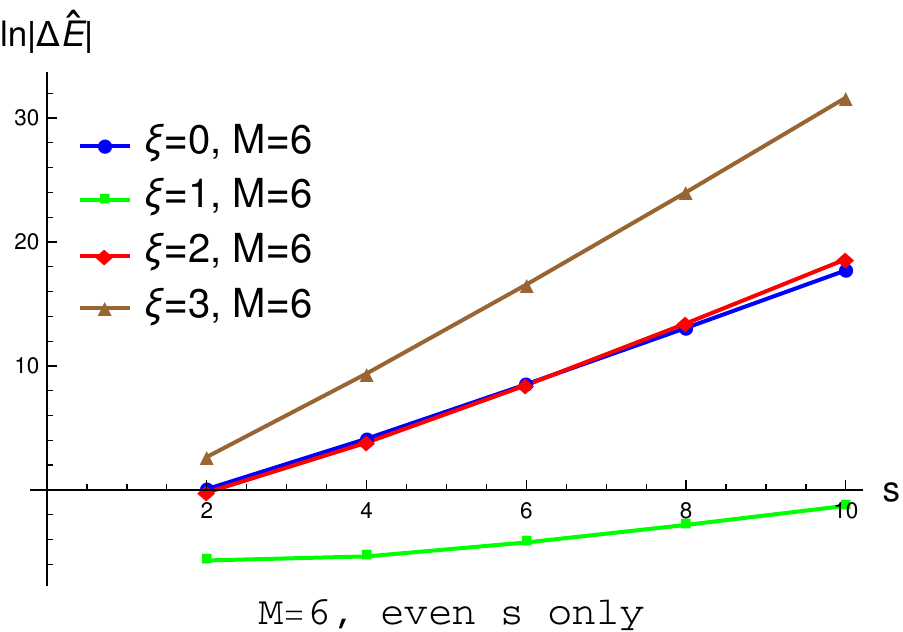}\includegraphics[width=0.5\textwidth]{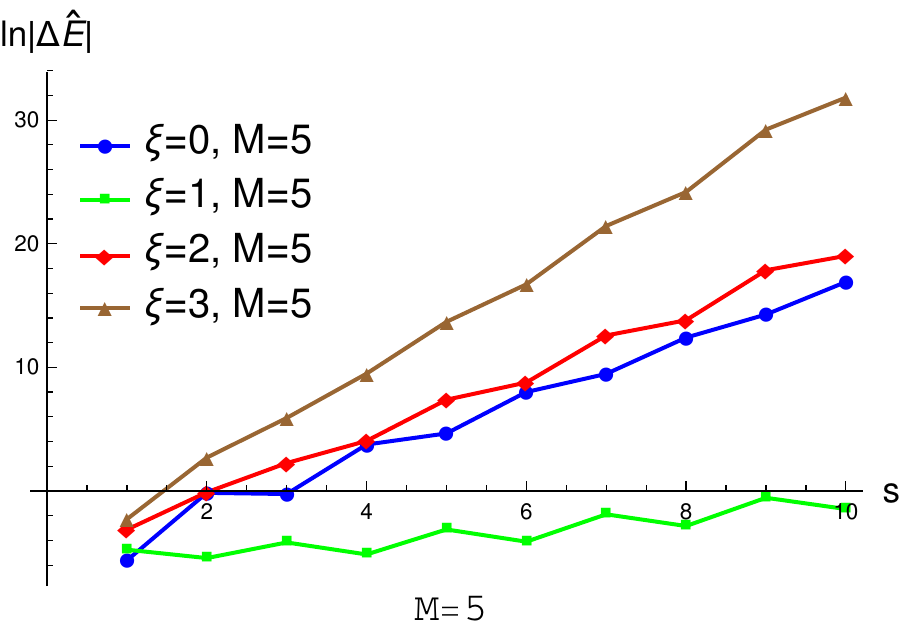}
\par\end{centering}
\caption{\label{fig:EvS}$\ln\left|\Delta\hat{E}\right|$ as a function of
$s$ for $M=6$ and $M=5$ cases. Note that for $M=6$ the $\xi=0$
(blue) and $\xi=2$ (red) curves almost coincide.}
\end{figure}

We next study the dependence of $\Delta\hat{E}=\sum_{i}\Delta\hat{E}_{i}$
on $s$. Figure \ref{fig:EvS} plots the change of $\ln\left|\Delta\hat{E}\right|$
with respect to $s$ for chains of $M=5$ and $M=6$. For $M=5$,
we sampled $s$ from 1 to $10$; for $M=6$, only even $s$ points
are sampled as its ground states only survive in even $s$ cases.
For each $M$, we choose $\xi=0,\,1,\,2,\,3$. For $M=6$, all the
curves almost rise linearly. Of all four curves, $\xi=3$ is the steepest
one, and $\xi=1$ is the flattest one. $\xi=0$ and $\xi=2$ almost
coincide with each other. For the $M=5$ case, the overall trends
of the curves are the same as $M=6$ except for slight oscillations
between even and odd $s$ points. For $\xi=0,\,1$, the oscillation
is relatively noticeable, and for $\xi=3$, it is negligible. Actually,
if only even $s$ points of $M=5$ are sampled, the plots are almost
the same as $M=6$. The exponential dependence of $\Delta\hat{E}$
on $s$ stems from the fact that each ground state has $2^{s}$ degeneracies.
The fact that $\xi=1$ has a lower slope than others is also related
to the fact that $\xi=1$ is the boundary for $\Delta E$ to be bounded
from below.

\section{\label{sec:Conclusion}Conclusion}

We have presented a formalism to calculate the cubic interaction vertices
for the stable string bit model. With the vertices, we calculated
the one-loop self-energies of the model in both analytical and numerical
ways. 

From the large $M$ behavior of one-loop self-energies, we found that
the Lorentz invariance requires the critical dimension of the model
to be $s=24$, which then leads to the protostring model. One interesting
interpretation of $s=24$ is as follows\cite{Thorn:2017pri}. Out
of the 24 dimensions, 16 of them are paired to form 8 compactified
bosonic dimensions, and the rest 8 remain as fermionic dimensions.
Thus, it has the same degrees of freedom as the superstring model.
The large $M$ behavior of $\Delta E_{G}$ is determined by the ground
states contribution of the small chains. Notwithstanding that the
number of excited states grows exponentially with respect to $M$
\cite{Chen:2016hkz}, the excited states contributions are canceled
out due to the fermionic nature of string bits. These results support
the idea of formulating string theory by string bit models.

The future research of this work can be done in several ways. One
can improve the numerical computation to study higher $M$ or $s$
cases. One can also apply the formalism to other calculations, e.g.,
four strings interaction, or to study higher-loop corrections and
find the Feynman rules of the model.

\section{Acknowledgments}

We thank Charles Thorn, Pierre Ramond, and Sourav Raha for helpful
discussions and comments on this work. This research was supported
in part by the Department of Energy under Grant No. DE-SC0010296.

\appendix
\numberwithin{equation}{section}

\section{\label{sec:app_hamiltonian}Hamiltonian and its action on color singlets}

The (anti)communication relations among string bit creation and annihilation
operators is
\begin{align}
\left[\left(\phi_{a_{1}\cdots a_{n}}\right)_{\alpha}^{\beta},\left(\bar{\phi}_{b_{1}\cdots b_{m}}\right)_{\gamma}^{\delta}\right]_{\pm} & \equiv\left(\phi_{a_{1}\cdots a_{n}}\right)_{\alpha}^{\beta}\left(\bar{\phi}_{b_{1}\cdots b_{m}}\right)_{\gamma}^{\delta}-\left(-1\right)^{mn}\left(\bar{\phi}_{b_{1}\cdots b_{m}}\right)_{\gamma}^{\delta}\left(\phi_{a_{1}\cdots a_{n}}\right)_{\alpha}^{\beta}\nonumber \\
 & =\delta_{mn}\delta_{\alpha}^{\delta}\delta_{\gamma}^{\beta}\sum_{P}\left(-1\right)^{P}\delta_{a_{1}b_{P_{1}}}\cdots\delta_{a_{n}b_{P_{n}}},\label{eq:App-H-commutation}
\end{align}
 where the sum runs over all permutations of $1,2,\dots,n$. 

The Hamiltonian of the model consists of $\mathcal{O}\left(1\right)$
terms and $\mathcal{O}\left(1/N\right)$ terms. The $\mathcal{O}\left(1\right)$
terms are the generalization of the Hamiltonian of the $s=1$ string
bit model \cite{Sun:2014dga,Chen:2016hkz} 
\begin{equation}
H^{s=1}=\frac{2}{N}\tr\left[\left(\bar{a}^{2}-i\bar{b}^{2}\right)a^{2}-\left(\bar{b}^{2}-i\bar{a}^{2}\right)b^{2}+\left(\bar{a}\bar{b}+\bar{b}\bar{a}\right)ba+\left(\bar{a}\bar{b}-\bar{b}\bar{a}\right)ab\right],\label{eq:App_ham_Hs_1}
\end{equation}
 where $\bar{a}=\bar{\phi}$ and $\bar{b}=\bar{\phi}_{1}$. $H^{s=1}$
produces the Green-Schwarz Hamiltonian \cite{Green:1980zg,Green:1983hw}
at $N=\infty$. 

$H^{s=1}$ is generalized to $\sum_{i=1}^{5}H_{i}$, where\cite{Thorn:2014hia,Thorn:2015wli}
\begin{subequations}\label{eq:App_ham_subh}
\begin{equation}
H_{1}=\frac{2}{N}\sum_{n=0}^{s}\sum_{k=0}^{s}\frac{s-2n}{n!k!}\tr\bar{\phi}_{a_{1}\cdots a_{n}}\bar{\phi}_{b_{1}\cdots b_{k}}\phi_{b_{1}\cdots b_{k}}\phi_{a_{1}\cdots a_{n}},\label{eq:App_ham_subh:1}
\end{equation}
\begin{equation}
H_{2}=\frac{2}{N}\sum_{n=0}^{s-1}\sum_{k=0}^{s-1}\frac{\left(-1\right)^{k}}{n!k!}\tr\bar{\phi}_{a_{1}\cdots a_{n}}\bar{\phi}_{bb_{1}\cdots b_{k}}\phi_{b_{1}\cdots b_{k}}\phi_{ba_{1}\cdots a_{n}},\label{eq:App_ham_subh:2}
\end{equation}
\begin{equation}
H_{3}=\frac{2}{N}\sum_{n=0}^{s-1}\sum_{k=0}^{s-1}\frac{\left(-1\right)^{k}}{n!k!}\tr\bar{\phi}_{ba_{1}\cdots a_{n}}\bar{\phi}_{b_{1}\cdots b_{k}}\phi_{bb_{1}\cdots b_{k}}\phi_{a_{1}\cdots a_{n}},\label{eq:App_ham_subh:3}
\end{equation}
\begin{equation}
H_{4}=\frac{2i}{N}\sum_{n=0}^{s-1}\sum_{k=0}^{s-1}\frac{\left(-1\right)^{k}}{n!k!}\tr\bar{\phi}_{a_{1}\cdots a_{n}}\bar{\phi}_{b_{1}\cdots b_{k}}\phi_{bb_{1}\cdots b_{k}}\phi_{ba_{1}\cdots a_{n}},\label{eq:App_ham_subh:4}
\end{equation}
\begin{equation}
H_{5}=-\frac{2i}{N}\sum_{n=0}^{s-1}\sum_{k=0}^{s-1}\frac{\left(-1\right)^{k}}{n!k!}\tr\bar{\phi}_{ba_{1}\cdots a_{n}}\bar{\phi}_{bb_{1}\cdots b_{k}}\phi_{b_{1}\cdots b_{k}}\phi_{a_{1}\cdots a_{n}}.\label{eq:App_ham_subh:5}
\end{equation}
\end{subequations}One can check that for $s=1$ eq. (\ref{eq:App_ham_subh})
is reduced to eq. (\ref{eq:App_ham_Hs_1}) if one identifies $\bar{\phi}$
as $\bar{a}$ and $\bar{\phi}_{1}$ as $\bar{b}$.

We now add $\mathcal{O}\left(1/N\right)$ terms to the Hamiltonian.
As refs. \cite{Chen:2016hkz,Sun:2014dga} show, the $N=\infty$ behavior
is not affected by the $\mathcal{O}\left(1/N\right)$ terms
\begin{align}
\Delta H^{s=1} & =\frac{2}{N}\tr\left[\bar{a}\bar{b}ba+\bar{b}\bar{a}ab+\bar{a}^{2}a^{2}+\bar{b}^{2}b^{2}-\tilde{M}^{s=1}\right],\\
\tilde{M}^{s=1} & =\tr\left(\bar{a}a+\bar{b}b\right)-\frac{1}{N}\left(\tr\bar{a}\tr a+\tr\bar{b}\tr b\right).
\end{align}
By analogy with $H^{s=1}$, $\Delta H^{s=1}$ can be generalized to
the arbitrary $s$ case as \begin{subequations}\label{eq:App_ham_deltah}
\begin{align}
\Delta H & =\frac{2}{N}\left(\sum_{n=0}^{s}\sum_{k=0}^{s}\frac{1}{n!k!}\tr\bar{\phi}_{b_{1}\cdots b_{k}}\bar{\phi}_{a_{1}\cdots a_{n}}\phi_{a_{1}\cdots a_{n}}\phi_{b_{1}\cdots b_{k}}-\tilde{M}\right),\\
\tilde{M} & =\sum_{n=0}^{s}\frac{1}{n!}\tr\bar{\phi}_{a_{1}\cdots a_{n}}\phi_{a_{1}\cdots a_{n}}-\frac{1}{N}\sum_{n=0}^{s}\frac{1}{n!}\tr\bar{\phi}_{a_{1}\cdots a_{n}}\tr\phi_{a_{1}\cdots a_{n}}.
\end{align}
\end{subequations}Combining the two parts together, we have the complete
form of the Hamiltonian for arbitrary $s$, 
\begin{equation}
H=\sum_{i=1}^{5}H_{i}+s\xi\Delta H
\end{equation}
where $\xi$ is a real constant. 

$H$ commutes with the supersymmetry operators 
\begin{equation}
Q^{a}=\sum_{n=0}^{s-1}\frac{\left(-1\right)^{n}}{n!}\tr\left[e^{i\pi/4}\bar{\phi}_{a_{1}\cdots a_{n}}\phi_{aa_{1}\cdots a_{n}}+e^{-i\pi/4}\bar{\phi}_{aa_{1}\cdots a_{n}}\phi_{a_{1}\cdots a_{n}}\right],\label{eq:App_ham_Q}
\end{equation}
\begin{equation}
\left\{ Q^{a},Q^{b}\right\} =2M\delta_{ab},
\end{equation}
 which will guarantee equal numbers of bosonic and fermionic eigenstates
at each energy level. 

Using the commutation relations (\ref{eq:App-H-commutation}), we
obtain the action of $H_{i}$ on single trace states \cite{Thorn:2015wli}
\begin{align*}
H_{1}T\left(\theta_{1},\cdots,\theta_{M}\right)\ket{0} & =2\sum_{k=1}^{M}\left(s-2\theta_{k}^{a}\frac{d}{d\theta_{k}^{a}}\right)T\left(\theta_{1},\cdots,\theta_{M}\right)\ket{0}\\
 & \hspace{1em}+\frac{2}{N}\sum_{k=1}^{M}\left(s-2\theta_{k}^{a}\frac{d}{d\theta_{k}^{a}}\right)\sum_{l\neq k,k+1}T\left(\theta_{1},\cdots,\theta_{k}\right)T\left(\theta_{k+1},\cdots,\theta_{l-1}\right)\ket{0},
\end{align*}
 
\begin{align*}
H_{2}T\left(\theta_{1},\cdots,\theta_{M}\right)\ket{0} & =2\sum_{k=1}^{M}\theta_{k}^{a}\frac{d}{d\theta_{k+1}^{a}}T\left(\theta_{1},\cdots,\theta_{M}\right)\ket{0}\\
 & \hspace{1em}+\frac{2}{N}\sum_{k=1}^{M}\sum_{l\neq k,k+1}\theta_{k}^{a}\frac{d}{d\theta_{l}^{a}}T\left(\theta_{1},\cdots,\theta_{k}\right)T\left(\theta_{k+1},\cdots,\theta_{l-1}\right)\ket{0},
\end{align*}
\begin{align*}
H_{3}T\left(\theta_{1},\cdots,\theta_{M}\right)\ket{0} & =2\sum_{k=1}^{M}\theta_{k+1}^{a}\frac{d}{d\theta_{k}^{a}}T\left(\theta_{1},\cdots,\theta_{M}\right)\ket{0}\\
 & \hspace{1em}+\frac{2}{N}\sum_{k=1}^{M}\sum_{l\neq k,k+1}\theta_{l}^{a}\frac{d}{d\theta_{k}^{a}}T\left(\theta_{1},\cdots,\theta_{k}\right)T\left(\theta_{k+1},\cdots,\theta_{l-1}\right)\ket{0},
\end{align*}
\begin{align*}
H_{4}T\left(\theta_{1},\cdots,\theta_{M}\right)\ket{0} & =-2i\sum_{k=1}^{M}\theta_{k}^{a}\theta_{k+1}^{a}T\left(\theta_{1},\cdots,\theta_{M}\right)\ket{0}\\
 & \hspace{1em}-\frac{2i}{N}\sum_{k=1}^{M}\sum_{l\neq k,k+1}\theta_{k}^{a}\theta_{l}^{a}T\left(\theta_{1},\cdots,\theta_{k}\right)T\left(\theta_{k+1},\cdots,\theta_{l-1}\right)\ket{0},
\end{align*}
\begin{align*}
H_{5}T\left(\theta_{1},\cdots,\theta_{M}\right)\ket{0} & =-2i\sum_{k=1}^{M}\frac{d}{d\theta_{k}^{a}}\frac{d}{d\theta_{k+1}^{a}}T\left(\theta_{1},\cdots,\theta_{M}\right)\ket{0}\\
 & \hspace{1em}-\frac{2i}{N}\sum_{k=1}^{M}\sum_{l\neq k,k+1}\frac{d}{d\theta_{k}^{a}}\frac{d}{d\theta_{l}^{a}}T\left(\theta_{1},\cdots,\theta_{k}\right)T\left(\theta_{k+1},\cdots,\theta_{l-1}\right)\ket{0}.
\end{align*}
 Similarly, the action of $\Delta H$ on a single trace state is
\[
\Delta HT\left(\theta_{1},\dots,\theta_{M}\right)\ket{0}=\frac{2}{N}\sum_{i=1}^{M}\sum_{j\neq i+1}^{M}T\left(\theta_{j},\dots\theta_{i}\right)T\left(\theta_{i+1},\dots\theta_{j-1}\right)\ket{0}.
\]
 The actions of $H_{i}$ on double traces are \cite{Thorn:2015wli}
\begin{align*}
H_{1}T\left(\theta_{1}\cdots\theta_{K}\right)T\left(\eta_{1}\cdots\eta_{L}\right)\ket{0}_{\mathrm{Fussion}} & =\frac{2}{N}\sum_{k=1}^{K}\sum_{l=1}^{L}\left(s-2\theta_{k}^{a}\frac{d}{d\theta_{k}^{a}}\right)T\left(\theta_{k+1}\cdots\theta_{k}\eta_{l}\cdots\eta_{l-1}\right)\ket{0}\\
 & \hspace{1em}+\frac{2}{N}\sum_{k=1}^{K}\sum_{l=1}^{L}\left(s-2\eta_{l}^{a}\frac{d}{d\eta_{l}^{a}}\right)T\left(\theta_{k}\cdots\theta_{k-1}\eta_{l+1}\cdots\eta_{l}\right)\ket{0},
\end{align*}
\begin{align*}
H_{2}T\left(\theta_{1}\cdots\theta_{K}\right)T\left(\eta_{1}\cdots\eta_{L}\right)\ket{0}_{\mathrm{Fussion}} & =\frac{2}{N}\sum_{k=1}^{K}\sum_{l=1}^{L}\theta_{k}^{a}\frac{d}{d\eta_{l}}T\left(\theta_{k+1}\cdots\theta_{k}\eta_{l}\cdots\eta_{l-1}\right)\ket{0}\\
 & \hspace{1em}+\frac{2}{N}\sum_{k=1}^{K}\sum_{l=1}^{L}\eta_{l}^{a}\frac{d}{d\theta_{k}^{a}}T\left(\theta_{k}\cdots\theta_{k-1}\eta_{l+1}\cdots\eta_{l}\right)\ket{0}.
\end{align*}
 Similarly, the action of $\Delta H$ on double traces is
\begin{align*}
\Delta HT\left(\theta_{1}\cdots\theta_{K}\right)T\left(\eta_{1}\cdots\eta_{L}\right)\ket{0}_{\mathrm{Fussion}} & =\frac{2}{N}\sum_{k=1}^{K}\sum_{l=1}^{L}T\left(\theta_{k+1}\cdots\theta_{k}\eta_{l}\cdots\eta_{l-1}\right)\ket{0}\\
 & \hspace{1em}+\frac{2}{N}\sum_{k=1}^{K}\sum_{l=1}^{L}T\left(\theta_{k}\cdots\theta_{k-1}\eta_{l+1}\cdots\eta_{l}\right)\ket{0}.
\end{align*}

\subsection{\label{subsec:DeltaH}Derivation of $\Delta H$}

It is not obvious how to generalize $\Delta H^{s=1}$ to arbitrary
$s$ cases. We actually obtain the generalization from the relation
\[
\tr G^{2}=N\left(\Delta H-H^{\prime}\right),
\]
 which has been proven in Appendix E of ref. \cite{Chen:2016hkz}
for $s=1$. Here, the color operator $G$ is defined as \cite{Bergman:1995wh}
\[
G_{\alpha}^{\beta}=\sum_{n=0}^{s}\frac{1}{n!}\left(\bar{\phi}_{a_{1}\cdots a_{n}}\phi_{a_{1}\cdots a_{n}}-\normord{\phi_{a_{1}\cdots a_{n}}\bar{\phi}_{a_{1}\cdots a_{n}}}\right)_{\alpha}^{\beta},
\]
 and both $\Delta H$ and $H^{\prime}$ are supersymmetric and of
$\mathcal{O}\left(1/N\right)$. The notation$\normord{\phi_{a_{1}\cdots a_{n}}\bar{\phi}_{a_{1}\cdots a_{n}}}$
indicates the normal ordering of $\phi_{a_{1}\cdots a_{n}}\bar{\phi}_{a_{1}\cdots a_{n}}$.
In $s=1$, we have\cite{Chen:2016hkz} 
\[
H^{\prime s=1}=\frac{2}{N}\tr\left(\bar{a}\normord{a\bar{a}}a+\bar{b}\normord{b\bar{a}}a-\bar{a}\normord{b\bar{b}}a\right).
\]
One can verify that the action of $G_{\alpha}^{\beta}$ on any color
singlet vanishes: $G_{\alpha}^{\beta}\ket{\text{any color singlet}}=0$.
We therefore have $\left(\Delta H-H^{\prime}\right)=0$ in the color
singlet space. 

To find $\Delta H$, we expand $\tr G^{2}$ and match its terms with
$H^{\prime s=1}$ and $\Delta H^{s=1}$. By direct calculation, we
have 
\begin{align*}
\tr G^{2} & =\sum_{n=0}^{s}\sum_{k=0}^{s}\frac{1}{n!k!}\tr\bar{\phi}_{a_{1}\cdots a_{n}}\phi_{a_{1}\cdots a_{n}}\bar{\phi}_{b_{1}\cdots b_{k}}\phi_{b_{1}\cdots b_{k}}+\sum_{n=0}^{s}\sum_{k=0}^{s}\frac{1}{n!k!}\tr\normord{\phi_{a_{1}\cdots a_{n}}\bar{\phi}_{a_{1}\cdots a_{n}}}\normord{\phi_{b_{1}\cdots a_{k}}\bar{\phi}_{b_{1}\cdots b_{k}}}\\
 & -\sum_{n=0}^{s}\sum_{k=0}^{s}\frac{1}{n!k!}\tr\bar{\phi}_{a_{1}\cdots a_{n}}\phi_{a_{1}\cdots a_{n}}\normord{\phi_{b_{1}\cdots b_{k}}\bar{\phi}_{b_{1}\cdots b_{k}}}-\sum_{n=0}^{s}\sum_{k=0}^{s}\frac{1}{n!k!}\tr\normord{\phi_{a_{1}\cdots a_{n}}\bar{\phi}_{a_{1}\cdots a_{n}}}\bar{\phi}_{b_{1}\cdots b_{k}}\phi_{b_{1}\cdots b_{k}}.
\end{align*}
 We calculate each term on the rhs of $\tr G^{2}$ and obtain 
\[
\text{first term}=N\sum_{n=0}^{s}\sum_{k=0}^{s}\frac{1}{n!}\tr\bar{\phi}_{a_{1}\cdots a_{n}}\phi_{a_{1}\cdots a_{n}}+\sum_{n=0}^{s}\sum_{k=0}^{s}\frac{\left(-1\right)^{nk}}{n!k!}\tr\bar{\phi}_{a_{1}\cdots a_{n}}\normord{\phi_{a_{1}\cdots a_{n}}\bar{\phi}_{b_{1}\cdots b_{k}}}\phi_{b_{1}\cdots b_{k}},
\]
\[
\text{second term}=N\sum_{n=0}^{s}\frac{1}{n!}\tr\bar{\phi}_{a_{1}\cdots a_{n}}\phi_{a_{1}\cdots a_{n}}+\sum_{n=0}^{s}\sum_{k=0}^{s}\frac{1}{n!k!}\tr\bar{\phi}_{a_{1}\cdots a_{n}}\normord{\phi_{b_{1}\cdots b_{k}}\bar{\phi}_{b_{1}\cdots b_{k}}}\phi_{a_{1}\cdots a_{n}},
\]
\[
\text{third term}=-\sum_{n=0}^{s}\frac{1}{n!}\tr\bar{\phi}_{a_{1}\cdots a_{n}}\tr\phi_{a_{1}\cdots a_{n}}-\sum_{n=0}^{s}\sum_{k=0}^{s}\frac{1}{n!k!}\tr\bar{\phi}_{b_{1}\cdots b_{k}}\bar{\phi}_{a_{1}\cdots a_{n}}\phi_{a_{1}\cdots a_{n}}\phi_{b_{1}\cdots b_{k}},
\]
\[
\text{fourth term}=-\sum_{n=0}^{s}\frac{1}{n!}\tr\bar{\phi}_{a_{1}\cdots a_{n}}\tr\phi_{a_{1}\cdots a_{n}}-\sum_{n=0}^{s}\sum_{k=0}^{s}\frac{1}{n!k!}\tr\bar{\phi}_{a_{1}\cdots a_{n}}\bar{\phi}_{b_{1}\cdots b_{k}}\phi_{b_{1}\cdots b_{k}}\phi_{a_{1}\cdots a_{n}}.
\]
 Combining the above together, we have
\begin{align*}
\tr G^{2} & =\sum_{n=0}^{s}\sum_{k=0}^{s}\frac{\left(-1\right)^{nk}}{n!k!}\tr\bar{\phi}_{a_{1}\cdots a_{n}}\normord{\phi_{a_{1}\cdots a_{n}}\bar{\phi}_{b_{1}\cdots b_{k}}}\phi_{b_{1}\cdots b_{k}}\\
 & +\sum_{n=0}^{s}\sum_{k=0}^{s}\frac{1}{n!k!}\tr\bar{\phi}_{a_{1}\cdots a_{n}}\normord{\phi_{b_{1}\cdots b_{k}}\bar{\phi}_{b_{1}\cdots b_{k}}}\phi_{a_{1}\cdots a_{n}}\\
 & -\sum_{n=0}^{s}\sum_{k=0}^{s}\frac{2}{n!k!}\tr\bar{\phi}_{b_{1}\cdots b_{k}}\bar{\phi}_{a_{1}\cdots a_{n}}\phi_{a_{1}\cdots a_{n}}\phi_{b_{1}\cdots b_{k}}\\
 & +N\sum_{n=0}^{s}\frac{2}{n!}\tr\bar{\phi}_{a_{1}\cdots a_{n}}\phi_{a_{1}\cdots a_{n}}-\sum_{n=0}^{s}\frac{2}{n!}\tr\bar{\phi}_{a_{1}\cdots a_{n}}\tr\phi_{a_{1}\cdots a_{n}}.
\end{align*}

Comparing the terms of $\tr G^{2}$ with $H^{\prime s=1}$ and $\Delta H^{s=1}$,
we can identify
\[
H^{\prime}=\frac{1}{N}\sum_{n=0}^{s}\sum_{k=0}^{s}\frac{1}{n!k!}\left(\tr\bar{\phi}_{a_{1}\cdots a_{n}}\normord{\phi_{b_{1}\cdots b_{k}}\bar{\phi}_{b_{1}\cdots b_{k}}}\phi_{a_{1}\cdots a_{n}}+\left(-\right)^{nk}\tr\bar{\phi}_{a_{1}\cdots a_{n}}\normord{\phi_{a_{1}\cdots a_{n}}\bar{\phi}_{b_{1}\cdots b_{k}}}\phi_{b_{1}\cdots b_{k}}\right),
\]
 and $\Delta H$ as (\ref{eq:App_ham_deltah}). One can verify that
both $H^{\prime}$ and $\Delta H$ commute with the supersymmetry
operators $Q^{a}$ (\ref{eq:App_ham_Q}). 

\section{\label{sec:App_normal_psibar}Verifying the normalization condition
for $\bar{\psi}_{G}$}

In this Appendix we show that the conjugate eigenfunction of $s=1$,
defined as eq. (\ref{eq:conj_psibarG}), satisfies the normalization
condition (\ref{eq:conj_normalization}). We first show that $\int d^{M}\theta\,\bar{\psi}_{G}\psi_{G}=1$.
For odd $M$, 
\begin{align*}
\int d^{M}\theta\,\bar{\psi}_{G}\psi_{G} & =\left(-i\right)^{\left\lfloor M/2\right\rfloor }\int d^{M}\theta\,\alpha_{0}\prod_{i=1}^{\left\lfloor M/2\right\rfloor }\left(-s_{i}+c_{i}\alpha_{M-i}\alpha_{i}\right)\left(c_{i}-s_{i}\alpha_{M-i}\alpha_{i}\right)\\
 & =\left(-i\right)^{\left\lfloor M/2\right\rfloor }\int d^{M}\theta\,\alpha_{0}\prod_{i=1}^{\left\lfloor M/2\right\rfloor }\left(s_{i}^{2}+c_{i}^{2}\right)\alpha_{M-i}\alpha_{i}\\
 & =\left(-i\right)^{\left\lfloor M/2\right\rfloor }\int d^{M}\theta\,\prod_{i=1}^{M}\alpha_{M-i}\\
 & =1,
\end{align*}
 where in the last step we used\footnote{We do not prove the formula (\ref{eq:App_psiG_alpha_prod}) here.
But we have verified it by the \textit{Mathematica} program.}
\begin{equation}
\int d^{M}\theta\,\prod_{i=1}^{M}\alpha_{M-i}\equiv\int d^{M}\theta\,\alpha_{M-1}\alpha_{M-2}\cdots\alpha_{0}=i^{\left\lfloor \left(M-1\right)/2\right\rfloor }.\label{eq:App_psiG_alpha_prod}
\end{equation}
Similarly, we can show $\int d^{M}\theta\,\bar{\psi}_{G}\psi_{G}=1$
for even $M$. 

To show$\int d^{M}\theta\,\bar{\psi}_{G}\psi_{r}=0$ for $r\neq G$,
it suffices to show that $\int d^{M}\theta\,\bar{\psi}_{G}F_{k}^{\dagger}\psi^{\prime}$
vanishes for all $0\leq k\leq M-1$ and any eigenfunction $\psi^{\prime}$.
If $k=0$, it clearly vanishes because both $F_{0}^{\dagger}$ and
$\bar{\psi}_{G}$ contain the Grassmann odd operator $\alpha_{0}$.
If $0<k<M/2$, 
\begin{equation}
\int d^{M}\theta\,\bar{\psi}_{G}F_{k}^{\dagger}\psi^{\prime}=\int d^{M}\theta\,\left(\tilde{F}_{k}^{\dagger\pm}\bar{\psi}_{G}\right)\psi^{\prime}.\label{eq:App_psiG_norm1}
\end{equation}
The rhs of (\ref{eq:App_psiG_norm1}) vanishes because of 
\begin{equation}
\tilde{F}_{k}^{\dagger\pm}\bar{\psi}_{G}=0,\label{eq:App_psiG_FtildeG}
\end{equation}
which can be verified by checking that 
\[
\tilde{F}_{k}^{\dagger\pm}\left(-s_{k}+c_{k}\alpha_{M-k}\alpha_{k}\right)=\tilde{F}_{M-k}^{\dagger\pm}\left(-s_{k}+c_{k}\alpha_{M-k}\alpha_{k}\right)=0,
\]
\[
\left[\tilde{F}_{k}^{\dagger\pm},-s_{l}+c_{l}\alpha_{M-l}\alpha_{l}\right]=0,\quad k\neq l,\,k\neq M-l.
\]
Similarly, we can show that $\tilde{F}_{k}^{\dagger\pm}\bar{\psi}_{G}=0$
for $M/2\leq k\leq M-1$. Therefore, the normalization condition $\int d^{M}\theta\,\bar{\psi}_{G}\psi_{r}=\delta_{Gr}$
is proved. 

\section{\label{sec:App_h_kl}Calculation of $h_{kl}$}

In this Appendix, we will find the expression of $h_{kl}^{a}$ in
terms of lowering and raising operators. The $h_{kl}^{a}$ in the
language of $\theta_{k}$ is

\[
h_{kl}=-2\left(1-2\theta_{k}\frac{d}{d\theta_{k}}\right)-2\theta_{k}\frac{d}{d\theta_{l}}-2\theta_{l}\frac{d}{d\theta_{k}}-2i\theta_{k}\theta_{l}-2i\frac{d}{d\theta_{k}}\frac{d}{d\theta_{l}}+2\xi+2\delta_{k,l}.
\]
We now temporarily drop the last two constant terms and will add them
back in the end of the calculation.

Using (\ref{eq:dia_fourier:2}), we express $\theta_{k}$ and $\frac{d}{d\theta_{k}}$
in terms of $\alpha_{n}$ and $\beta_{n}$:
\begin{align*}
\theta_{k}\frac{d}{d\theta_{k}} & =\frac{1}{M}\sum_{n,m=0}^{M-1}\alpha_{n}\beta_{m}\exp\left(2\pi ik\frac{m+n}{M}\right)\\
\theta_{k}\frac{d}{d\theta_{l}}+\theta_{l}\frac{d}{d\theta_{k}} & =\frac{1}{M}\sum_{n,m=0}^{M-1}\left(\alpha_{n}\beta_{m}+\alpha_{m}\beta_{n}\right)\left[\exp\left(2\pi i\frac{kn+lm}{M}\right)\right]\\
\theta_{k}\theta_{l} & =\frac{1}{M}\sum_{n,m=0}^{M-1}\alpha_{n}\alpha_{m}\exp\left(2\pi i\frac{kn+lm}{M}\right)\\
\frac{d}{d\theta_{k}}\frac{d}{d\theta_{l}} & =\frac{1}{M}\sum_{n,m=0}^{M-1}\beta_{n}\beta_{m}\exp\left(2\pi i\frac{kn+lm}{M}\right).
\end{align*}
 Substituting the above into $h_{kl}$ and rearranging, we obtain
\[
h_{kl}=h_{kl}^{\left(1\right)}+h_{kl}^{\left(0\right)}
\]
where $h_{kl}^{\left(0\right)}$ are the terms with zero modes and
$h_{kl}^{\left(1\right)}$ are the terms without, 
\begin{align*}
h_{kl}^{\left(1\right)} & =-2+\frac{4}{M}\sum_{n,m=1}^{M-1}\alpha_{n}\beta_{m}\exp\left(2\pi ik\frac{m+n}{M}\right)\\
 & \hspace{1em}-\frac{2}{M}\sum_{n,m=1}^{M-1}\left(\alpha_{n}\beta_{m}+\alpha_{m}\beta_{n}+i\alpha_{n}\alpha_{m}+i\beta_{n}\beta_{m}\right)\exp\left(2\pi i\frac{kn+lm}{M}\right),
\end{align*}
\[
h_{kl}^{\left(0\right)}=\frac{2}{M}\sum_{n=1}^{M-1}\left(\alpha_{n}\beta_{0}+\alpha_{0}\beta_{n}-i\alpha_{n}\alpha_{0}-i\beta_{n}\beta_{0}\right)\left[\exp\left(2\pi i\frac{kn}{M}\right)-\exp\left(2\pi i\frac{ln}{M}\right)\right].
\]

Let us first consider $h_{kl}^{\left(1\right)}$. We express nonzero
modes $\alpha_{m}$ and $\beta_{m}$ in terms of raising and lowering
operators. Using 
\[
\alpha_{k}=c_{k}F_{M-k}^{\dagger}+s_{k}F_{k},\quad\beta_{k}=-s_{k}F_{M-k}^{\dagger}+c_{k}F_{k},\quad k=1,\cdots,M-1,
\]
we have\begin{subequations}\label{eq:App_hkl_ab} 
\begin{equation}
\alpha_{n}\alpha_{m}=c_{n}c_{m}F_{M-n}^{\dagger}F_{M-m}^{\dagger}+s_{n}s_{m}F_{n}F_{m}+c_{n}s_{m}F_{M-n}^{\dagger}F_{m}-c_{m}s_{n}F_{M-m}^{\dagger}F_{n}+c_{n}s_{n}\delta_{m+n,M},\label{eq:App_hkl_ab:1}
\end{equation}
\begin{equation}
\alpha_{n}\beta_{m}=-c_{n}s_{m}F_{M-n}^{\dagger}F_{M-m}^{\dagger}+c_{m}s_{n}F_{n}F_{m}+c_{n}c_{m}F_{M-n}^{\dagger}F_{m}+s_{n}s_{m}F_{M-m}^{\dagger}F_{n}+s_{n}^{2}\delta_{m+n,M},\label{eq:App_hkl_ab:2}
\end{equation}
\begin{equation}
\beta_{n}\beta_{m}=s_{n}s_{m}F_{M-n}^{\dagger}F_{M-m}^{\dagger}+c_{n}c_{m}F_{n}F_{m}-c_{m}s_{n}F_{M-n}^{\dagger}F_{m}+c_{n}s_{m}F_{M-m}^{\dagger}F_{n}+c_{n}s_{n}\delta_{m+n,M}.\label{eq:App_hkl_ab:3}
\end{equation}
\end{subequations}We then apply eqs. (\ref{eq:App_hkl_ab}) to $h_{kl}^{\left(1\right)}$,
collect like terms, and antisymmetrize $F^{\dagger}F^{\dagger}$ and
$FF$ terms to give 
\begin{equation}
h_{kl}^{\left(1\right)}=\frac{2}{M}\sum_{n,m=1}^{M-1}\left(A_{nm}^{\dagger}F_{n}^{\dagger}F_{m}^{\dagger}+A_{nm}F_{n}F_{m}+2A_{-n,m}F_{n}^{\dagger}F_{m}\right)-\frac{2}{M}\cot\frac{\pi}{2M}+\frac{2}{M}\cot\frac{2\left(k-l\right)+1}{2M}\pi,\label{eq:App_hkl1}
\end{equation}
 where 
\begin{equation}
A_{nm}=\exp\left(2\pi ik\frac{m+n}{M}\right)\sin\frac{m-n}{2M}\pi+\frac{i}{2}\left[\exp\left(2\pi i\frac{km+ln}{M}\right)\exp\left(\pi i\frac{m-n}{2M}\right)-m\leftrightarrow n\right].\label{eq:App_hkl_Anm}
\end{equation}

Similarly, applying 
\[
\alpha_{n}\beta_{0}=\exp\left(-\frac{i\pi}{4}\right)c_{n}F_{M-n}^{\dagger}F_{0}+s_{n}F_{n}F_{0}
\]
\[
\alpha_{0}\beta_{n}=\exp\left(\frac{i\pi}{4}\right)\left(s_{n}F_{M-n}^{\dagger}F_{0}^{\dagger}+c_{n}F_{0}^{\dagger}F_{n}\right)
\]
\[
\alpha_{n}\alpha_{0}=\exp\left(\frac{i\pi}{4}\right)\left(c_{n}F_{M-n}^{\dagger}F_{0}^{\dagger}-s_{n}F_{0}^{\dagger}F_{n}\right)
\]
\[
\beta_{n}\beta_{0}=\exp\left(-\frac{i\pi}{4}\right)\left(-s_{n}F_{M-n}^{\dagger}F_{0}+c_{n}F_{n}F_{0}\right)
\]
 to $h_{kl}^{\left(0\right)}$ yields

\begin{align*}
h_{kl}^{\left(0\right)} & =\frac{2}{M}\sum_{n=1}^{M-1}\left(\alpha_{n}\beta_{0}+\alpha_{0}\beta_{n}-i\alpha_{n}\alpha_{0}-i\beta_{n}\beta_{0}\right)\left[\exp\left(2\pi i\frac{kn}{M}\right)-\exp\left(2\pi i\frac{ln}{M}\right)\right]\\
 & =\frac{2}{M}\sum_{n=1}^{M-1}X_{n}^{\dagger}\left(F_{0}^{\dagger}F_{n}^{\dagger}-F_{n}^{\dagger}F_{0}\right)+\mathrm{h.c.},
\end{align*}
 where
\[
X_{n}=i\left[\exp\left(2\pi i\frac{ln}{M}\right)-\exp\left(2\pi i\frac{kn}{M}\right)\right]\exp\left(-\frac{in\pi}{2M}\right).
\]
 Now from eq. (\ref{eq:App_hkl_Anm}), we see that 
\begin{align*}
A_{n0} & =-\exp\left(2\pi i\frac{kn}{M}\right)\sin\frac{n\pi}{2M}+\frac{i}{2}\left[\exp\left(2\pi i\frac{ln}{M}\right)\exp\left(-\frac{n\pi i}{2M}\right)-\exp\left(2\pi i\frac{kn}{M}\right)\exp\left(\frac{n\pi i}{2M}\right)\right]\\
 & =-\frac{i}{2}\exp\left(2\pi i\frac{kn}{M}\right)\exp\left(-\frac{in\pi}{2M}\right)+\frac{i}{2}\exp\left(2\pi i\frac{ln}{M}\right)\exp\left(-\frac{in\pi}{2M}\right)\\
 & =\frac{1}{2}X_{n}.
\end{align*}
Hence, to add $h_{kl}^{\left(0\right)}$ terms to $h_{kl}^{\left(1\right)}$,
we can simply change the $m,n$ index of (\ref{eq:App_hkl1}) to start
from $0$. Finally, adding back the constant terms, we have 
\begin{equation}
h_{kl}=\frac{2}{M}\sum_{n,m=0}^{M-1}\left(A_{nm}^{\dagger}F_{n}^{\dagger}F_{m}^{\dagger}+A_{nm}F_{n}F_{m}+2A_{-n,m}F_{n}^{\dagger}F_{m}\right)+\frac{2}{M}\mu\label{eq:App_hkl_h}
\end{equation}
 where $A_{mn}$ is given by (\ref{eq:App_hkl_Anm}) and 
\begin{equation}
\mu=-\cot\frac{\pi}{2M}+\cot\frac{2\left(k-l\right)+1}{2M}\pi+M\xi.\label{eq:App_hkl_mu}
\end{equation}

\section{\label{sec:app_AFF}Calculation of $\left\langle \cdots\bar{A}_{nm}F_{n}^{\dagger}F_{m}^{\dagger}\exp\left(\frac{1}{2}f_{k}^{\dagger}D_{kl}f_{l}^{\dagger}\right)\right\rangle $}

In this Appendix we will derive the formula 
\begin{equation}
\left\langle \cdots\bar{A}_{nm}F_{n}^{\dagger}F_{m}^{\dagger}\exp\left(\frac{1}{2}f_{k}^{\dagger}D_{kl}f_{l}^{\dagger}\right)\right\rangle =-\frac{2}{M}\left[\tr\left(S^{*}C^{-1}\bar{A}\right)+B_{mn}\frac{\partial}{\partial D_{mn}}\right]\left\langle \cdots\exp\left(\frac{1}{2}f_{k}^{\dagger}D_{kl}f_{l}^{\dagger}\right)\right\rangle \label{eq:App_AFF_1}
\end{equation}
 where $\bar{A}\equiv A^{\dagger}$ and $B=C^{-1}\bar{A}\left(C^{-1}\right)^{T}$,
and the relations among $F_{n}^{\dagger}$, $f_{k}$, and $f_{k}^{\dagger}$
are given by 
\begin{equation}
F_{m}=\sum_{n=0}^{M-1}\left(f_{n}C_{mn}+f_{n}^{\dagger}S_{mn}\right),\quad0\leq m\leq M-1.\label{eq:App_AFF_Ffrelation}
\end{equation}
 with $C_{mn}D_{nl}+S_{ml}=0$.

Let $X=\frac{1}{2}f_{k}^{\dagger}D_{kl}f_{l}^{\dagger}$, $\ket{G}=\exp\left(X\right)\ket{0}$;
then 
\begin{equation}
F_{n}^{\dagger}\bar{A}_{nm}F_{m}^{\dagger}\ket{G}=\exp\left(X\right)\left(\bar{A}_{nm}F_{n}^{\dagger}F_{m}^{\dagger}-\left[X,\bar{A}_{nm}F_{n}^{\dagger}F_{m}^{\dagger}\right]+\frac{1}{2}\left[X,\left[X,\bar{A}_{nm}F_{n}^{\dagger}F_{m}^{\dagger}\right]\right]+\cdots\right)\ket{0}.\label{eq:App_AFF_res1}
\end{equation}
Now let us calculate each term in the parentheses of the rhs of eq.
(\ref{eq:App_AFF_res1}). For the first term\begin{subequations}\label{eq:App_AFF_subres}
\begin{align}
\bar{A}_{nm}F_{n}^{\dagger}F_{m}^{\dagger}\ket{0} & =\left(\bar{A}_{nm}f_{i}^{\dagger}f_{j}^{\dagger}C_{ni}^{*}C_{mj}^{*}+\bar{A}_{nm}f_{i}S_{ni}^{*}f_{j}^{\dagger}C_{mj}^{*}\right)\ket{0}\nonumber \\
 & =\left(\bar{A}_{nm}f_{i}^{\dagger}f_{j}^{\dagger}C_{ni}^{*}C_{mj}^{*}+\bar{A}_{nm}S_{ni}^{*}C_{mj}^{*}\left\{ f_{i},f_{j}^{\dagger}\right\} \right)\ket{0}\nonumber \\
 & =\left[f_{i}^{\dagger}\left(C^{\dagger}\bar{A}C^{*}\right)_{ij}f_{j}^{\dagger}+\tr\left(\bar{A}C^{*}S^{\dagger}\right)\right]\ket{0}\label{eq:App_AFF_subres:1}
\end{align}
For the second term of the rhs of eq. (\ref{eq:App_AFF_res1}), we
first find 
\begin{align*}
\left[\frac{1}{2}f_{k}^{\dagger}D_{kl}f_{l}^{\dagger},F_{m}^{\dagger}\right] & =\frac{1}{2}D_{kl}\left[f_{k}^{\dagger}f_{l}^{\dagger},F_{m}^{\dagger}\right]\\
 & =\frac{1}{2}D_{kl}\left(f_{k}^{\dagger}\left\{ f_{l}^{\dagger},F_{m}^{\dagger}\right\} -\left\{ f_{k}^{\dagger},F_{m}^{\dagger}\right\} f_{l}^{\dagger}\right)\\
 & =\frac{1}{2}D_{kl}\left(f_{k}^{\dagger}\delta_{ln}S_{mn}^{*}-\delta_{kn}S_{mn}^{*}f_{l}^{\dagger}\right)\\
 & =-S_{ml}^{*}D_{lk}f_{k}^{\dagger},
\end{align*}
 where in the second step we used the identity $\left[AB,C\right]=A\left\{ B,C\right\} -\left\{ A,C\right\} B$
and in the last step we used the property that $D_{kl}$ is antisymmetric.
We then have 
\begin{align*}
\left[\frac{1}{2}f_{k}^{\dagger}D_{kl}f_{l}^{\dagger},F_{n}^{\dagger}\bar{A}_{nm}F_{m}^{\dagger}\right] & =F_{n}^{\dagger}\bar{A}_{nm}\left[\frac{1}{2}f_{k}^{\dagger}D_{kl}f_{l}^{\dagger},F_{m}^{\dagger}\right]+\left[\frac{1}{2}f_{k}^{\dagger}D_{kl}f_{l}^{\dagger},F_{n}^{\dagger}\right]\bar{A}_{nm}F_{m}^{\dagger}\\
 & =-F_{n}^{\dagger}\bar{A}_{nm}S_{ml}^{*}D_{lk}f_{k}^{\dagger}-S_{nl}^{*}D_{lk}f_{k}^{\dagger}\bar{A}_{nm}F_{m}^{\dagger}\\
 & =\bar{A}_{nm}S_{ml}^{*}D_{lk}\left(f_{k}^{\dagger}F_{n}^{\dagger}-\left\{ F_{n}^{\dagger},f_{k}^{\dagger}\right\} \right)-S_{nl}^{*}D_{lk}f_{k}^{\dagger}\bar{A}_{nm}F_{m}^{\dagger}\\
 & =\bar{A}_{nm}S_{ml}^{*}D_{lk}\left(f_{k}^{\dagger}F_{n}^{\dagger}-S_{nk}^{*}\right)-S_{nl}^{*}D_{lk}f_{k}^{\dagger}\bar{A}_{nm}F_{m}^{\dagger}\\
 & =\bar{A}_{nm}S_{ml}^{*}D_{lk}f_{k}^{\dagger}F_{n}^{\dagger}-S_{nl}^{*}D_{lk}f_{k}^{\dagger}\bar{A}_{nm}F_{m}^{\dagger}-\bar{A}_{nm}S_{ml}^{*}D_{lk}S_{nk}^{*}\\
 & =f_{k}^{\dagger}D_{kl}\left(S^{\dagger}\right)_{lm}\bar{A}_{mn}F_{n}^{\dagger}+f_{k}^{\dagger}D_{kl}\left(S^{\dagger}\right)_{ln}\bar{A}_{nm}F_{m}^{\dagger}-\bar{A}_{nm}S_{ml}^{*}D_{lk}\left(S^{\dagger}\right)_{kn}\\
 & =2f_{k}^{\dagger}\left(DS^{\dagger}\bar{A}\right)_{km}F_{m}^{\dagger}-\tr\left(\bar{A}S^{*}DS^{\dagger}\right).
\end{align*}
 It then follows that
\begin{equation}
\left[X,F_{n}^{\dagger}\bar{A}_{nm}F_{m}^{\dagger}\right]\ket{0}=\left[2f_{k}^{\dagger}\left(DS^{\dagger}\bar{A}C^{*}\right)_{kl}f_{l}^{\dagger}-\tr\left(\bar{A}S^{*}DS^{\dagger}\right)\right]\ket{0}.\label{eq:App_AFF_subres:2}
\end{equation}
 For the third term of the rhs of eq. (\ref{eq:App_AFF_res1}),
\begin{align}
\left[X,\left[X,\bar{A}_{nm}F_{n}^{\dagger}F_{m}^{\dagger}\right]\right] & =\left[X,2f_{k}^{\dagger}\left(DS^{\dagger}\bar{A}\right)_{km}F_{m}^{\dagger}\right]-\zero{\left[X,\tr\left(\bar{A}S^{*}DS^{\dagger}\right)\right]}\nonumber \\
 & =2f_{i}^{\dagger}\left(DS^{\dagger}\bar{A}\right)_{im}\left[\frac{1}{2}\sum_{k,l}f_{k}^{\dagger}D_{kl}f_{l}^{\dagger},F_{m}^{\dagger}\right]\nonumber \\
 & =-2f_{k}^{\dagger}\left(DS^{\dagger}\bar{A}S^{*}D\right)_{kl}f_{l}^{\dagger}.\label{eq:App_AFF_subres:3}
\end{align}
\end{subequations} It follows that the higher order commutations
all vanish. Substituting eqs. (\ref{eq:App_AFF_subres}) into eq.
(\ref{eq:App_AFF_res1}), we have
\begin{align*}
F_{n}^{\dagger}\bar{A}_{nm}F_{m}^{\dagger}\ket{G} & =\left[f_{k}^{\dagger}\left(C^{\dagger}\bar{A}C^{*}\right)_{kl}f_{l}^{\dagger}+\tr\left(\bar{A}C^{*}S^{\dagger}\right)\right]\ket{G}\\
 & \hspace{1em}-\left[2f_{k}^{\dagger}\left(DS^{\dagger}\bar{A}C^{*}\right)_{kl}f_{l}^{\dagger}-\tr\left(\bar{A}S^{*}DS^{\dagger}\right)\right]\ket{G}-f_{k}^{\dagger}\left(DS^{\dagger}\bar{A}S^{*}D\right)_{kl}f_{l}^{\dagger}\ket{G}\\
 & =\tr\left[\bar{A}\left(C^{*}+S^{*}D\right)S^{\dagger}\right]\ket{G}\\
 & \hspace{1em}+f_{k}^{\dagger}\left(C^{\dagger}\bar{A}C^{*}-2DS^{\dagger}\bar{A}C^{*}-DS^{\dagger}\bar{A}S^{*}D\right)_{kl}f_{l}^{\dagger}\ket{G}\\
 & =\tr\left[\bar{A}\left(C^{*}+S^{*}D\right)S^{\dagger}\right]\ket{G}\\
 & \hspace{1em}+f_{k}^{\dagger}\left(C^{\dagger}\bar{A}C^{*}-DS^{\dagger}\bar{A}C^{*}+C^{\dagger}\bar{A}S^{*}D-DS^{\dagger}\bar{A}S^{*}D\right)_{kl}f_{l}^{\dagger}\ket{G}\\
 & =\tr\left[\bar{A}\left(C^{*}+S^{*}D\right)S^{\dagger}\right]\ket{G}\\
 & \hspace{1em}+f_{k}^{\dagger}\left[C^{\dagger}\bar{A}\left(C^{*}+S^{*}D\right)-DS^{\dagger}\bar{A}\left(S^{*}D+C^{*}\right)\right]_{kl}f_{l}^{\dagger}\ket{G}\\
 & =\tr\left[\bar{A}\left(C^{*}+S^{*}D\right)S^{\dagger}\right]\ket{G}\\
 & \hspace{1em}+f_{k}^{\dagger}\left[\left(C^{*}+S^{*}D\right)^{T}\bar{A}\left(C^{*}+S^{*}D\right)\right]_{kl}f_{l}^{\dagger}\ket{G},
\end{align*}
 where in the third equality we antisymmetrized the $2DS^{\dagger}\bar{A}C^{*}$
term to be $DS^{\dagger}\bar{A}C^{*}-\left(DS^{\dagger}\bar{A}C^{*}\right)^{T}$
and then used the fact that $\bar{A}$ and $D$ are antisymmetric
matrices. Now, 
\[
C^{*}+S^{*}D=C^{*}-C^{*}D^{*}D=C^{*}\left(I-D^{*}D\right)=C^{*}\left(I+DD^{\dagger}\right)^{T}=C^{*}\left(C^{-1}C^{-1\dagger}\right)^{T}=\left(C^{-1}\right)^{T},
\]
 where in the second-to-last equality $I+DD^{\dagger}=C^{-1}C^{-1\dagger}$
follows from eqs. (\ref{eq:overlap_CS_relation}) and (\ref{eq:overlap_CMS}).
We therefore have 
\[
F_{n}^{\dagger}\bar{A}_{nm}F_{m}^{\dagger}\ket{G}=-\tr\left(S^{*}C^{-1}\bar{A}\right)\ket{G}+f_{k}^{\dagger}\left[C^{-1}\bar{A}\left(C^{-1}\right)^{T}\right]_{kl}f_{l}^{\dagger}\ket{G},
\]
 which implies (\ref{eq:App_AFF_1}).

\section{\label{sec:app_example}Examples of $M=3$}

In this Appendix, as a demo of using (\ref{eq:olp_DeltaE}) to calculate
one-loop self-energy, let us consider the one-loop self-energy for
the ground state of the $M=3,\,s=1$ and $M=3,\,s=2$ cases. For $M=3$,
we only need to calculate the $K=1$ case since the contribution of
$K=2$ is the same as $K=1$. 

The $C$, $S$, and $D$ matrices are 
\begin{eqnarray*}
C & = & \begin{pmatrix}1\end{pmatrix}\oplus\frac{1+\sqrt{3}}{4}\begin{pmatrix}e^{i\frac{\pi}{6}} & -e^{i\frac{2\pi}{3}}\\
e^{-i\frac{\pi}{6}} & e^{i\frac{2\pi}{3}}
\end{pmatrix},\quad\left|\det C\right|=\frac{2+\sqrt{3}}{4}\\
S & = & \begin{pmatrix}0\end{pmatrix}\oplus\frac{\sqrt{3}-1}{4}\begin{pmatrix}e^{-i\frac{\pi}{6}} & e^{-i\frac{2\pi}{3}}\\
e^{-i\frac{5\pi}{6}} & e^{-i\frac{\pi}{3}}
\end{pmatrix},\quad D=\begin{pmatrix}0\end{pmatrix}\oplus\left(2-\sqrt{3}\right)\begin{pmatrix}0 & i\\
-i & 0
\end{pmatrix},
\end{eqnarray*}
 and matrices $A$, $B$, and constant $\mu^{\prime}$ are 
\begin{eqnarray*}
A^{\left(V\right)} & = & \begin{pmatrix}0 & i\frac{\sqrt{3}}{4} & \frac{\sqrt{3}}{4}\\
-i\frac{\sqrt{3}}{4} & 0 & \frac{3}{4}\\
-\frac{\sqrt{3}}{4} & -\frac{3}{4} & 0
\end{pmatrix},\quad A^{\left(W\right)}=\begin{pmatrix}0 & \frac{i\sqrt{3}}{2} & \frac{\sqrt{3}}{2}\\
-\frac{i\sqrt{3}}{2} & 0 & 0\\
-\frac{\sqrt{3}}{2} & 0 & 0
\end{pmatrix}\\
B^{\left(V\right)} & = & \begin{pmatrix}0 & \frac{1}{2}\sqrt{\frac{3}{2}}e^{3i\pi/4} & \sqrt{\frac{3}{2}}\left(1-\frac{\sqrt{3}}{2}\right)e^{3i\pi/4}\\
\frac{1}{2}\sqrt{\frac{3}{2}}e^{-i\pi/4} & 0 & 3i\left(2-\sqrt{3}\right)\\
\sqrt{\frac{3}{2}}\left(1-\frac{\sqrt{3}}{2}\right)e^{-i\pi/4} & -3i\left(2-\sqrt{3}\right) & 0
\end{pmatrix},\\
B^{\left(W\right)} & = & \begin{pmatrix}0 & \sqrt{\frac{3}{2}}e^{3i\pi/4} & \sqrt{\frac{3}{2}}\left(2-\sqrt{3}\right)e^{3i\pi/4}\\
\sqrt{\frac{3}{2}}e^{-i\pi/4} & 0 & 0\\
\sqrt{\frac{3}{2}}\left(2-\sqrt{3}\right)e^{-i\pi/4} & 0 & 0
\end{pmatrix},\\
\mu^{\prime\left(V\right)} & = & -3+\sqrt{3}+3\xi,\quad\mu^{\prime\left(W\right)}=-4\sqrt{3}+6\xi.
\end{eqnarray*}
The operators $f_{n}$ are 
\[
f_{0}=F_{0},\quad f_{1}=F_{1}^{\left(2\right)},\quad f_{2}=e^{-i\pi/4}\left(\sqrt{\frac{2}{3}}F_{0}^{\left(1\right)}-F_{0}^{\left(2\right)}\sqrt{\frac{1}{3}}\right).
\]

For $s=1$, the eigenfunctions and their conjugates of $1$-bit and
$2$-bit chains are shown in Table \ref{tab:Energy-eigenstates-1}.

\begin{center}
\begin{table}
\begin{centering}
\begin{tabular}{|c|c|c|c|c|}
\hline 
$M$ & $\psi$ & Conjugate & Energy & Grading of $\psi$\tabularnewline
\hline 
\hline 
1 & $\psi_{G}^{\left(1\right)}=1$ & $\bar{\psi}_{G}^{\left(1\right)}=\theta_{1}$ & $E_{G}^{\left(1\right)}=0$ & even\tabularnewline
\hline 
2 & $\psi_{1}^{\left(2\right)}=F_{1}^{\left(2\right)\dagger}\psi_{G}^{\left(2\right)}$ & $\bar{\psi}_{1}^{\left(2\right)}=\tilde{F}_{1}^{\left(2\right)}\bar{\psi}_{G}^{\left(2\right)}$ & $E_{1}^{\left(2\right)}=4$ & odd\tabularnewline
\hline 
\end{tabular}
\par\end{centering}
\caption{\label{tab:Energy-eigenstates-1}1-bit and 2-bit energy eigenstates
of $s=1$ that do not contain zero modes.}
\end{table}
\par\end{center}

The contribution of $K=1$ to the energy correction is 
\begin{equation}
\Delta E_{G}^{K=1}=\frac{1}{N^{2}}\frac{KLM\left|\det C\right|}{E_{G}-E_{G}^{\left(1\right)}-E_{1}^{\left(2\right)}}\left(\left\langle F_{1}^{\left(2\right)}f_{0}\right\rangle _{W}^{*}\left\langle F_{1}^{\left(2\right)}f_{0}\right\rangle _{V}+\left\langle F_{1}^{\left(2\right)}f_{2}\right\rangle _{W}^{*}\left\langle F_{1}^{\left(2\right)}f_{2}\right\rangle _{V}\right).\label{eq:App_ex_deltaEp}
\end{equation}
So, we need to calculate $\left\langle F_{1}^{\left(2\right)}f_{0}\right\rangle _{V,W}$
and $\left\langle F_{1}^{\left(2\right)}f_{1}\right\rangle _{V,W}$:
\begin{align*}
\left\langle F_{1}^{\left(2\right)}f_{0}\right\rangle _{V} & =\frac{2}{M}\mu^{\prime\left(V\right)}\zero{\left\langle f_{1}f_{0}\exp\left(\frac{1}{2}f_{k}^{\dagger}D_{kl}f_{l}^{\dagger}\right)\right\rangle }+\frac{2}{M}B_{mn}^{\left(V\right)}\left\langle f_{1}f_{0}f_{m}^{\dagger}f_{n}^{\dagger}\exp\left(\frac{1}{2}f_{k}^{\dagger}D_{kl}f_{l}^{\dagger}\right)\right\rangle \\
 & =\frac{4}{M}B_{0n}^{\left(V\right)}\left\langle f_{1}f_{n}^{\dagger}\exp\left(\frac{1}{2}f_{k}^{\dagger}D_{kl}f_{l}^{\dagger}\right)\right\rangle \\
 & =\frac{4}{M}B_{01}^{\left(V\right)}=\sqrt{\frac{2}{3}}e^{3i\pi/4},
\end{align*}
\begin{align*}
\left\langle F_{1}^{\left(2\right)}f_{2}\right\rangle _{V} & =\frac{2}{M}\mu^{\prime\left(V\right)}\left\langle f_{1}f_{2}\exp\left(\frac{1}{2}f_{k}^{\dagger}D_{kl}f_{l}^{\dagger}\right)\right\rangle +\frac{2}{M}B_{mn}^{\left(V\right)}\left\langle f_{1}f_{2}f_{m}^{\dagger}f_{n}^{\dagger}\exp\left(\frac{1}{2}f_{k}^{\dagger}D_{kl}f_{l}^{\dagger}\right)\right\rangle \\
 & =-\frac{2}{M}\mu^{\prime\left(V\right)}D_{12}-\frac{4}{M}B_{12}^{\left(V\right)}=-2i\left(1-\frac{1}{\sqrt{3}}+2\xi-\sqrt{3}\xi\right).
\end{align*}
 Likewise,
\begin{align*}
\left\langle F_{1}^{\left(2\right)}f_{0}\right\rangle _{W} & =\frac{4}{M}B_{01}^{\left(W\right)}=2\sqrt{\frac{2}{3}}e^{3i\pi/4},\\
\left\langle F_{1}^{\left(2\right)}f_{2}\right\rangle _{W} & =-\frac{2}{M}\mu^{\prime\left(W\right)}D_{12}-\frac{4}{M}B_{12}^{\left(W\right)}=-8i\left(1-\frac{2}{\sqrt{3}}+\xi-\frac{\sqrt{3}}{2}\xi\right).
\end{align*}
Substituting above results and $E_{G}=-4\sqrt{3}$ into (\ref{eq:App_ex_deltaEp})
yields $\Delta E_{G}^{K=1}=-\frac{3}{2}\left(3\sqrt{3}-5\right)\xi^{2}+\left(12-7\sqrt{3}\right)\xi-\frac{3}{4}\left(3\sqrt{3}-5\right)$.
We then have 
\[
\Delta E_{G}=2\Delta E_{G}^{L=1}=-3\left(3\sqrt{3}-5\right)\xi^{2}+2\left(12-7\sqrt{3}\right)\xi-\frac{3}{2}\left(3\sqrt{3}-5\right).
\]

For $s=2$, the matrices and constants are the same as the $s=1$
case. But as Table \ref{tab:Energy-eigenstates-2} shows, the energy
eigenstates of small chains are different. The energy correction now
is given by 
\begin{align}
\Delta E_{G}^{K=1} & =\frac{1}{N^{2}}\frac{KLM\left|\det C\right|^{2}}{E_{G}-E_{G}^{\left(2\right)}-0}\left(\left\langle 1\right\rangle _{W}^{*}\left\langle 1\right\rangle _{V}+\left\langle f_{2}f_{0}\right\rangle _{W}^{*}\left\langle f_{2}f_{0}\right\rangle _{V}\right)^{2}\nonumber \\
 & \hspace{1em}+\frac{1}{N^{2}}\frac{KLM\left|\det C\right|^{2}}{E_{G}-E_{1}^{\left(2\right)}-0}\left(\left\langle F_{1}^{\left(2\right)}f_{0}\right\rangle _{W}^{*}\left\langle F_{1}^{\left(2\right)}f_{0}\right\rangle _{V}+\left\langle F_{1}^{\left(2\right)}f_{2}\right\rangle _{W}^{*}\left\langle F_{1}^{\left(2\right)}f_{2}\right\rangle _{V}\right)^{2}.\label{eq:App_ex_DeltaEs2}
\end{align}
Since we have calculated the $\left\langle F_{1}^{\left(2\right)}f_{0}\right\rangle _{V,W}$
and $\left\langle F_{1}^{\left(2\right)}f_{2}\right\rangle _{V,W}$
in the $s=1$ case, we only need to find $\left\langle 1\right\rangle _{V,W}$
and $\left\langle f_{2}f_{0}\right\rangle _{V,W}$. 

\begin{center}
\begin{table}
\begin{centering}
\begin{tabular}{|c|c|c|c|c|}
\hline 
$M$ & $\psi$ & Conjugate & Energy & Grading of $\psi$\tabularnewline
\hline 
\hline 
1 & $\psi_{G}^{\left(1\right)}=1$ & $\bar{\psi}_{G}^{\left(1\right)}=\theta_{1}$ & $E_{G}^{\left(1\right)}=0$ & even\tabularnewline
\hline 
2 & $\psi_{G}^{\left(2\right)}$ & $\bar{\psi}_{G}^{\left(2\right)}$ & $E_{G}^{\left(2\right)}=-8$ & even\tabularnewline
\hline 
2 & $\psi_{1}^{\left(2\right)}=F_{1,a=1}^{\left(2\right)\dagger}F_{1,a=2}^{\left(2\right)\dagger}\psi_{G}^{\left(2\right)}$ & $\bar{\psi}_{1}^{\left(2\right)}=\tilde{F}_{1,a=2}^{\left(2\right)}\tilde{F}_{1,a=1}^{\left(2\right)}\bar{\psi}_{G}^{\left(2\right)}$ & $E_{1}^{\left(2\right)}=8$ & even\tabularnewline
\hline 
\end{tabular}
\par\end{centering}
\caption{\label{tab:Energy-eigenstates-2}1-bit and 2-bit energy eigenstates
of $s=2$ that do not contain zero modes.}
\end{table}
\par\end{center}

For $K=1$,

\begin{align*}
\left\langle 1\right\rangle _{V} & =\frac{2}{M}\mu^{\prime\left(V\right)}\left\langle \exp\left(\frac{1}{2}f_{k}^{\dagger}D_{kl}f_{l}^{\dagger}\right)\right\rangle +\frac{2}{M}B_{mn}^{\left(V\right)}\zero{\left\langle f_{m}^{\dagger}f_{n}^{\dagger}\exp\left(\frac{1}{2}f_{k}^{\dagger}D_{kl}f_{l}^{\dagger}\right)\right\rangle }\\
 & =\frac{2}{M}\mu^{\prime\left(V\right)}=\frac{2}{\sqrt{3}}-2+2\xi,
\end{align*}
\begin{align*}
\left\langle f_{2}f_{0}\right\rangle _{V} & =\frac{2}{M}\mu^{\prime\left(V\right)}\zero{\left\langle f_{2}f_{0}\exp\left(\frac{1}{2}f_{k}^{\dagger}D_{kl}f_{l}^{\dagger}\right)\right\rangle }+\frac{2}{M}B_{mn}^{\left(V\right)}\left\langle f_{2}f_{0}f_{m}^{\dagger}f_{n}^{\dagger}\exp\left(\frac{1}{2}f_{k}^{\dagger}D_{kl}f_{l}^{\dagger}\right)\right\rangle \\
 & =\frac{4}{M}B_{02}^{\left(V\right)}=\sqrt{2}\left(\frac{2}{\sqrt{3}}-1\right)e^{3i\pi/4},
\end{align*}
 Likewise
\[
\left\langle 1\right\rangle _{W}=\frac{2}{M}\gamma^{\prime\left(W\right)}=-\frac{8}{\sqrt{3}}+4\xi,\quad\left\langle f_{2}f_{0}\right\rangle _{W}=\frac{4}{M}B_{02}^{\left(W\right)}=2\sqrt{2}\left(\frac{2}{\sqrt{3}}-1\right)e^{3i\pi/4}.
\]
 Substituting the above into eq. (\ref{eq:App_ex_DeltaEs2}), we obtain
\[
\Delta E_{G}=2\Delta E_{G}^{K=1}=\frac{1}{N^{2}}\left(-66\sqrt{3}\xi^{4}+360\xi^{3}-230\sqrt{3}\xi^{2}+180\xi-\frac{33\sqrt{3}}{2}\right).
\]

From the results and the formula (\ref{eq:olp_DeltaE}), we see that
$\Delta E_{G}$ is a polynomial of $\xi$ of degree $2s$.

\bibliographystyle{plain}
\bibliography{ref}

\begin{thebibliography}{10}

\bibitem{Thorn:1991fv}
C.~B. Thorn.
\newblock {Reformulating string theory with the 1/N expansion}.
\newblock In {\em {The First International A.D. Sakharov Conference on Physics
  Moscow, USSR, May 27-31, 1991}}, 1991.

\bibitem{'tHooft:1987}
G.~'t~Hooft.
\newblock {On the Quantization of Space and Time}.
\newblock In V.A.~Berezin Eds. M.A.~Markov and V.P. Frolov, editors, {\em
  {{Proc. of the 4th Seminar on Quantum Gravity}, May 25-29, 1987, Moscow,
  USSR.}}, pages 551--567. World Scientific Press, 1988.

\bibitem{'tHooft:1990eb}
G.~'t~Hooft.
\newblock {Quantization of Discrete Deterministic Theories by Hilbert Space
  Extension}.
\newblock {\em Nucl. Phys.}, B342:471--485, 1990.

\bibitem{'tHooft:1993gx}
G.~'t~Hooft.
\newblock {Dimensional reduction in quantum gravity}.
\newblock In {\em {Salamfest 1993:0284-296}}, pages 0284--296, 1993.

\bibitem{'tHooft:1973jz}
G.~'t~Hooft.
\newblock {A Planar Diagram Theory for Strong Interactions}.
\newblock {\em Nucl. Phys.}, B72:461, 1974.

\bibitem{Thorn:1979gu}
C.~B. Thorn.
\newblock {A Fock Space Description of the 1/$N_c$ Expansion of Quantum
  Chromodynamics}.
\newblock {\em Phys. Rev.}, D20:1435, 1979.

\bibitem{Bergman:1995wh}
O.~Bergman and C.~B. Thorn.
\newblock {String bit models for superstring}.
\newblock {\em Phys. Rev.}, D52:5980--5996, 1995.

\bibitem{Sun:2014dga}
S.~Sun and C.~B. Thorn.
\newblock {Stable String Bit Models}.
\newblock {\em Phys. Rev.}, D89(10):105002, 2014.

\bibitem{Thorn:2014hia}
C.~B. Thorn.
\newblock {Space from String Bits}.
\newblock {\em JHEP}, 11:110, 2014.

\bibitem{Chen:2016hkz}
G.~Chen and S.~Sun.
\newblock {Numerical Study of the Simplest String Bit Model}.
\newblock {\em Phys. Rev.}, D93(10):106004, 2016.

\bibitem{Thorn:2015wli}
C.~B. Thorn.
\newblock {1/N Perturbations in Superstring Bit Models}.
\newblock {\em Phys. Rev.}, D93(6):066003, 2016.

\bibitem{Chen:2015GitHub}
G.~Chen.
\newblock {String bit project} source code.
\newblock \url{https://github.com/gaolichen/stringbit}.
\newblock Accessed: 2016-01-23.

\bibitem{Thorn:2017pri}
C.~B. Thorn.
\newblock Personal communication.

\bibitem{Green:1980zg}
M.~B. Green and J.~H. Schwarz.
\newblock {Supersymmetrical Dual String Theory}.
\newblock {\em Nucl. Phys.}, B181:502--530, 1981.

\bibitem{Green:1983hw}
M.~B. Green, J.~H. Schwarz, and L.~Brink.
\newblock {Superfield Theory of Type II Superstrings}.
\newblock {\em Nucl. Phys.}, B219:437--478, 1983.

\end{thebibliography}

\end{document}